\documentclass[lettersize,journal]{IEEEtran}
\usepackage{amsmath,amsfonts}
\usepackage{algorithmic}
\usepackage{array}
\usepackage[caption=false,font=normalsize,labelfont=sf,textfont=sf]{subfig}
\usepackage{textcomp}
\usepackage{stfloats}
\usepackage{url}
\usepackage{verbatim}
\usepackage{graphicx}
\usepackage{multirow}
\usepackage{cite}
\usepackage{amssymb}
\usepackage{amsthm,amsmath}
\usepackage{mathrsfs}
\usepackage{indentfirst}
\usepackage{multirow}
\usepackage{subcaption}
\usepackage[ruled,linesnumbered]{algorithm2e}
\usepackage{grffile}
\usepackage{makecell}
\usepackage[breaklinks,colorlinks,citecolor=blue]{hyperref}
\begin{document}

\title{Image Restoration via Multi-domain Learning}

\author{Xingyu Jiang, Ning Gao, Xiuhui Zhang, Hongkun Dou, Shaowen Fu, Xiaoqing Zhong, Hongjue Li and \\ Yue Deng,~\IEEEmembership{Senior Member,~IEEE,}
        % <-this % stops a space
\thanks{Xingyu Jiang, Ning Gao, Xiuhui Zhang, Hongkun Dou, Hongjue Li and Yue Deng are with Beihang University, Beijing 100191, China; Shaowen Fu is with the Beijing Aerospace Automatic Control Institute; Xiaoqing Zhong is with the China Academy of Space Technology. (e-mail: ydeng@buaa.edu.cn)}% <-this % stops a space
% \thanks{corresponding author: Yue Deng.}
}

\markboth{Journal of \LaTeX\ Class Files,~Vol.~14, No.~8, August~2021}%
{Shell \MakeLowercase{\textit{et al.}}: A Sample Article Using IEEEtran.cls for IEEE Journals}


\maketitle

\begin{abstract}
Due to adverse atmospheric and imaging conditions, natural images suffer from various degradation phenomena. Consequently, image restoration has emerged as a key solution and garnered substantial attention. Although recent Transformer architectures have demonstrated impressive success across various restoration tasks, their considerable model complexity poses significant challenges for both training and real-time deployment. Furthermore, instead of investigating the commonalities among different degradations, most existing restoration methods focus on modifying Transformer under limited restoration priors. In this work, we first review various degradation phenomena under multi-domain perspective, identifying common priors. Then, we introduce a novel restoration framework, which integrates multi-domain learning into Transformer. Specifically, in Token Mixer, we propose a Spatial-Wavelet-Fourier multi-domain structure that facilitates local-region-global multi-receptive field modeling to replace vanilla self-attention. Additionally, in Feed-Forward Network, we incorporate multi-scale learning to fuse multi-domain features at different resolutions. Comprehensive experimental results across ten restoration tasks, such as dehazing, desnowing, motion deblurring, defocus deblurring, rain streak/raindrop removal, cloud removal, shadow removal, underwater enhancement and low-light enhancement, demonstrate that our proposed model outperforms state-of-the-art methods and achieves a favorable trade-off among restoration performance, parameter size, computational cost and inference latency. The code is available at: https://github.com/deng-ai-lab/SWFormer.
\end{abstract}

\begin{IEEEkeywords}
Image Restoration, Image Enhancement, Global Modeling, Spatial-Wavelet-Fourier Multi-domain Learning.
\end{IEEEkeywords}

\section{Introduction}

\begin{figure*}[!t]
	\centering
	\includegraphics[width=\linewidth]{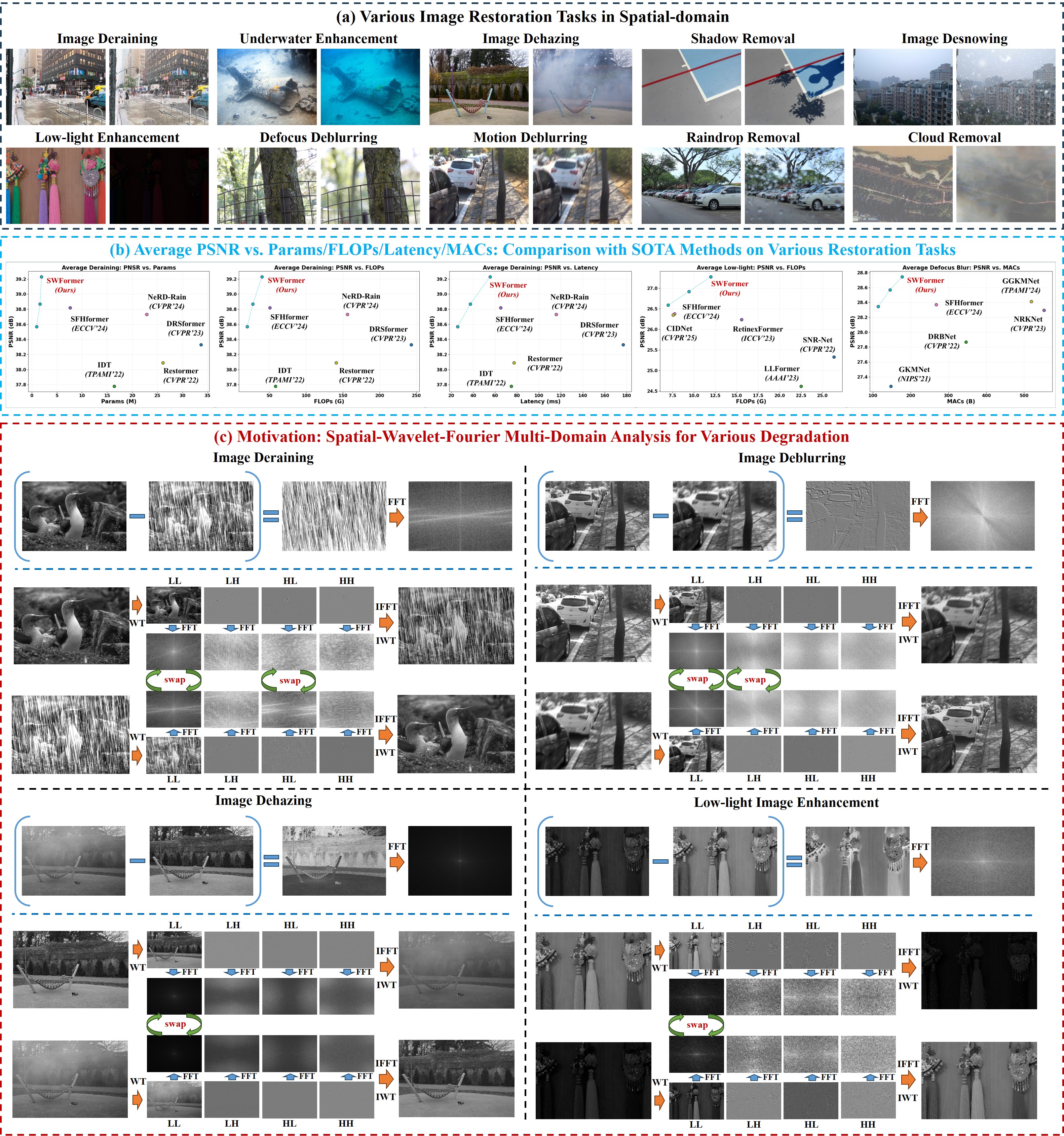}
	\caption{(a) Spatial-domain representations for various restoration tasks. (b) Performance and model complexity balance: Average PSNR vs. Params/FLOPs/Latency/MACs. We average performance across different benchmarks for statistical characteristics. (c) Motivation: Spatial-Wavelet-Fourier analysis for various degradations. For simplicity, we present the results in grayscale.}
    \vspace{-4mm}
	\label{fig:intro}
\end{figure*}

\IEEEPARstart{N}{atural} images can suffer from various degradation processes (as shown in Fig.\ref{fig:intro}(a)), resulting from atmospheric scattering effects (e.g., dehazing\cite{dcp}, remote sensing cloud removal\cite{emrdm}, underwater enhancement\cite{waternetuieb}), dynamic medium interference (e.g., rain streak/rain drop removal\cite{RAIN200, AttentiveGANraindrop}, snow removal\cite{desnownesnow100k}), optical system imperfections (e.g., defocus blur removal\cite{DPDD}, motion blur removal\cite{GOPRO}) and lighting condition limitations (e.g., low-light enhancement\cite{retinexformer}, shadow removal\cite{ShadowFormer}). These degradations lead to significant reductions in image contrast, color shift and geometric distortions, which severely affect image quality assessment and hinder the performance of downstream visual tasks\cite{tpamidetection, tpamidetection1}. To this end, image restoration has emerged as a key solution, with the goal of reconstructing high-quality clear images from low-quality degraded inputs, and has garnered extensive attention in the computer vision community over the past few decades. However, due to the inherently ill-posed nature of restoration problem, designing an efficient model capable of effectively handling various degradation tasks remains a highly complex challenge. So far, current image restoration techniques can generally be classified into two categories: prior-based methods\cite{dcp, SPDNet} and data-driven methods\cite{sfhformer, fadformer}. Prior-based models treat image restoration as an ill-posed optimization problem, incorporating physical priors to constrain the solution space, but may struggle in complex, real-world scenarios. In contrast, data-driven models, particularly convolutional neural networks (CNNs), employ an end-to-end manner to solve image restoration and have achieved successful breakthroughs. More recently, Transformer-like self-attention approaches\cite{Restormer, DRSformer} have emerged and delivered state-of-the-art (SOTA) performance across various tasks. 

Despite the significant progress made by Transformer-like approaches in image restoration, the accompanying increase in model complexity presents substantial challenges for both training and real-time deployment. To address this problem, existing restoration methods have attempted to optimize the Transformer architecture. For instance, Restormer\cite{Restormer} introduces channel self-attention, GRL\cite{GRL-B} incorporates anchor self-attention and DRSformer\cite{DRSformer} adopts sparse self-attention to reduce model complexity. However, these approaches overlook the common priors across different degradation processes in network design, failing to effectively integrate the inherent characteristics of restoration tasks themselves. With limited restoration priors, existing methods struggle to achieve an optimal balance among performance, parameter size, computational cost and inference latency. In Fig.\ref{fig:intro}(b), DRSformer\cite{DRSformer} approach has $\times$35 more parameters, $\times$12 more computational cost and $\times$6.8 longer inference latency than our SWFormer-s, yet its average PSNR score is 0.24dB lower than ours.

To this end, we begin by reviewing various degradation phenomena from Spatial-Wavelet-Fourier perspective and observe that multi-domain transformations can effectively separate degradation representations, providing valuable insights for restoration network design. In the following, we elaborate on the underlying discoveries behind multi-domain perspective. \textbf{(1) Compact Fourier Domain Degradation Representation:} In Fig.\ref{fig:intro}(c), the first row shows the residuals of clear and degraded images, alongside their Fourier-domain representations via Fast Fourier Transform (FFT). Degradation in the spatial domain is often globally diffuse, whereas in the Fourier domain, it is more compact and localized. Specifically, in tasks like deraining and deblurring, degradation primarily affects high frequencies, while in dehazing and low-light enhancement, it concentrates in low frequencies. This suggests that modeling degradation in the Fourier domain offers more efficient solutions. \textbf{(2) Strong Separation Restoration Priors:} In subsequent rows of Fig.\ref{fig:intro}(c), we apply Wavelet Transform (WT) to decompose the image into four sub-bands: [LL, LH, HL, HH], followed by FFT to map them into the Fourier domain. We find that degradation tends to concentrate in specific sub-bands, such as LL and HL for deraining, LL and LH for deblurring and LL for dehazing and low-light. We swap the specific sub-bands between the clear and degraded images, and through inverse transformations, recover the corresponding swapped versions. Consequently, the initial clear image transform into its degraded counterpart and conversely, the degraded image exhibits clear one. This demonstrates that WT and FFT allow us to isolate degradation representations into specific wavelet sub-bands at targeted Fourier frequencies, providing a strong restoration prior for network design. \textbf{(3) Efficient Local-Region-Global Receptive Field Modeling:} The Spatial-Wavelet-Fourier multi-domain perspective aligns with local-region-global receptive field modeling, where the spatial domain captures local features, the wavelet domain models regional features and the Fourier domain encodes global features. Additionally, the Fourier domain’s \(O(N \log N)\) computational complexity is more efficient than the \(O(N^2)\) complexity of self-attention. Building on these insights, we integrate these restoration priors into the transformer architecture, enabling efficient image restoration via multi-domain learning.

In this work, we propose SWFormer, a novel restoration backbone designed to tackle various degradation tasks from a multi-domain perspective. Inspired by NAFNet\cite{NAFNet}, SWFormer introduces key enhancements at both the inter-block and intra-block levels. At the inter-block level, we introduce a Lossless Multi-Input Multi-Output (LMIMO) framework, enabling dynamic output generation at different sizes (small, medium and large) and performance boosting with minimal computational overhead. At the intra-block level, we retain the Transformer architecture but modify the Token Mixer and Feed-Forward Network. Specifically, SWFormer comprises two essential components: the Spatial-Wavelet-Fourier Mixer (SWFM) and the Multi-scale ConvFFN (MSFN). The SWFM replaces self-attention\cite{visiontransformer} with a Spatial-Wavelet-Fourier tri-branch structure for local-region-global modeling, reducing model complexity. The wavelet branch adopts learnable convolutions to parametrize the wavelet decomposition, while the Fourier branch introduces a gating mechanism to filter frequency components. For the MSFN, we incorporate multi-scale learning to aggregate features from the spatial, wavelet and Fourier domains across different resolutions, enabling more comprehensive and efficient feature fusion.

We evaluate the effectiveness of SWFormer through extensive experiments on 26 widely adopted benchmarks across 10 restoration tasks, including dehazing\cite{griddehazenet}, desnowing\cite{JSTASRSRRS}, rain streak/raindrop removal\cite{SPA-Data, AttentiveGANraindrop}, cloud removal\cite{de-msda_memory}, motion blur\cite{GOPRO}, defocus blur\cite{DPDD}, shadow removal\cite{DHAN}, underwater enhancement\cite{U-shapeTranslsui} and low-light enhancement\cite{lolv1}. The results demonstrate that SWFormer achieves state-of-the-art (SOTA) performance on most datasets, while maintaining an excellent balance between restoration performance, parameter size, computational cost and inference latency.

We summarize the main contributions as follows:
\begin{itemize}
    \item[$\bullet$] We revisit various degradation phenomena from a Spatial-Wavelet-Fourier perspective, providing valuable restoration priors and insights for future network designs. 
    
    \item[$\bullet$] We propose SWFormer, a novel and efficient image restoration backbone that handles a wide range of restoration tasks, with a carefully designed architecture at both inter-block and intra-block levels.
    
    \item[$\bullet$] At inter-block level, we introduce a Lossless Multi-Input Multi-Output framework to enhance performance and enable dynamic outputs. At intra-block level, we integrate multi-domain learning into the Transformer for local-region-global modeling and incorporate multi-scale learning for feature aggregation.
    
    \item[$\bullet$] We demonstrate that SWFormer surpasses SOTA methods, maintaining favorable balance between restoration performance, parameter size, computational cost and inference latency across 10 restoration tasks.
\end{itemize}

Compared to its conference version \cite{sfhformer}, this present work expands on more detailed discussion, analysis and materials, including new ideas, network architecture designs, additional experimental results and more comprehensive ablation studies. \textbf{(1) New Discoveries:} We revisit the restoration priors for various degradation tasks from a Spatial-Wavelet-Fourier multi-domain perspective. Unlike the conference version, we introduce the wavelet domain for more precise separation of degradation across sub-bands, providing valuable insights for network architecture design. \textbf{(2) New Architecture Designs:} We introduce new network structures at both the inter-block and intra-block levels compared with the conference version. At the inter-block level, we update the original single-input single-output architecture into a lossless multi-input multi-output framework, enabling dynamic generation of models at varying sizes (small, medium and large) and showcasing the scaling capability of our model. Besides, such framework can improve restoration quality with minimal computational overhead. At the intra-block level, we add a wavelet branch in Token Mixer to extract region features and incorporate gating mechanism in the Fourier branch to replace the dynamic convolution, improving Fourier domain filtering efficiency. \textbf{(3) Additional Experiments:} Compared with the conference version, we further validate the effectiveness of SWFormer on two new restoration tasks: remote sensing cloud removal and shadow removal. Additionally, for defocus deblurring, we conduct experiments on a new larger-scale dataset LFDOF\cite{aifnet}. Extensive ablation studies confirm the effectiveness of the newly introduced components in our model.

\begin{figure*}[!b]
	\centering
	\includegraphics[width=\linewidth]{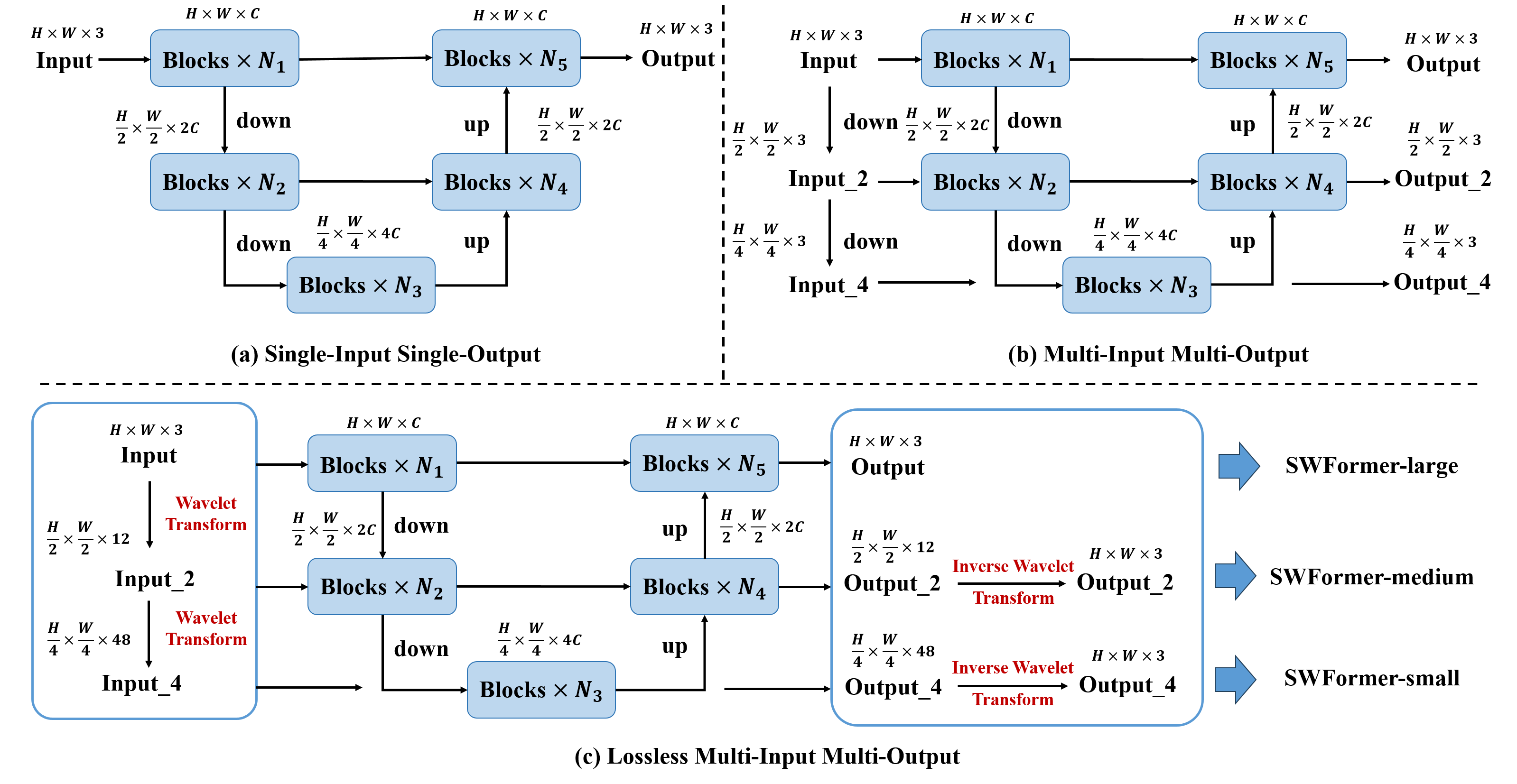}
	\caption{Inter-block designs: (a) Single-Input Single-Output, (b) Multi-Input Multi-Output, (c) Lossless Multi-Input Multi-Output.
	}
	\label{fig:inter-block}
\end{figure*}

\section{Related Work}
In this section, we first review the development of image restoration and then summarize the characteristics of current model architectures at both the inter-block and intra-block levels, highlighting the differences in our method.

\subsection{Image Restoration}
Image restoration is a long-standing ill-posed optimization problem that has drawn significant interest in the computer vision community. To handle this ill-posed nature, additional prior information or statistical constraints are often introduced. Current image restoration methods can be broadly classified into two categories: prior-based methods\cite{dcp, SPDNet, retinex} and data-driven methods\cite{MSBDN, Restormer, focalnet, Uformer, shadowdiffusion, jiang2022boosting}. Prior-based methods typically start from the restoration tasks themself, constructing degradation equations and introducing novel physical assumptions or statistical findings to constrain the solution space, thus solving the ill-posed optimization problem. For example, \cite{dcp} proposed the dark channel prior for dehazing, \cite{SPDNet} introduced the residual channel prior for deraining and \cite{retinex} proposed the Retinex prior model for low-light. These prior-based methods demonstrate nice statistical properties in certain scenarios, but may fail when the assumptions no longer hold in complex real-world scenes.

Recently, data-driven methods, particularly CNNs\cite{ffaNET, MPRNet}, have been proposed to overcome the limitations of prior-based approaches. Thanks to novel architectural designs (such as residual networks\cite{resnet}, feature pyramids\cite{pymaridfeature} and dense blocks\cite{densenet}) and unique mechanisms (such as attention\cite{ffaNET}, adversarial learning\cite{CycleGAN} and multi-kernel\cite{inception}), these CNN-based methods have achieved impressive performance in end-to-end manner. More recently, vision transformers\cite{visiontransformer} have introduced the self-attention mechanism into computer vision, gradually gaining popularity in the low-level vision community. Unlike CNNs, the self-attention captures long-range dependencies through global modeling. Various transformer-based algorithms\cite{Restormer, Uformer} have demonstrated the significance of global modeling in image restoration and have achieved remarkable success in tasks such as dehazing\cite{dehazeformer}, deraining\cite{DRSformer}, motion deblurring\cite{fftformer} and low-light enhancement\cite{retinexformer}. However, the substantial model complexity introduced by the self-attention poses significant challenges for both training and real-time deployment. As a result, numerous simplification methods\cite{Restormer, GRL-B, mambair} for self-attention have been proposed. For instance, Swin-Transformer\cite{liu2021swin} introduced window self-attention, Restormer\cite{Restormer} proposed channel self-attention and Mambair\cite{mambair} introduced visual state-space model as an efficient alternative to self-attention. Although these methods have been successful in reducing model complexity, they often come at the cost of decreased performance, making it difficult to strike an optimal balance between performance, model size and computational cost.

Unlike previous studies, we aim to combine the strengths of both prior-based and data-driven methods. By incorporating restoration priors into the design of data-driven model architectures, we propose a customized backbone specifically tailored for image restoration. Specifically, we first revisit the common priors across different degradation processes from spatial-wavelet-Fourier perspective and then leverage multi-domain learning to achieve local-region-global multi-receptive field modeling, enabling efficient image restoration.

\subsection{Architecture Design at Inter-block Level}
Fig.\ref{fig:inter-block} illustrates different designs of inter-block level architecture. Initially, most models\cite{fadformer, sfhformer} adopt a single-input single-output (SISO) architecture in Fig.\ref{fig:inter-block}(a), where the degraded image is input into the network at the beginning and a clear output is obtained at the end. Recently, multi-input multi-output (MIMO) architectures\cite{sfnet, IRNeXt} in Fig.\ref{fig:inter-block}(b), have gradually become mainstream in model design, in which multi-scale degraded images are input at different stages, with outputs at corresponding stages as well. Compared to SISO, MIMO enhances the extraction of multi-scale features and can be viewed as auxiliary learning. MIMO provides extra constraints to optimize the network at different stages.

In our work, as shown in Fig.\ref{fig:inter-block}(c), we further introduce the Lossless Multi-Input Multi-Output (LMIMO) architecture based on MIMO. Compared with MIMO, the LMIMO architecture features the following core improvements: (1) For the multi-input part, we employ Wavelet Transform or Pixel Unshuffle\cite{pixelshuffle} for downsampling, which preserves lossless information and enriches the multi-scale feature inputs. (2) For the multi-output part, we utilize Inverse Wavelet Transform or Pixel Shuffle\cite{pixelshuffle} for upsampling, allowing the network to generate restored images at different stages. This design further enables dynamic forward inference within the network. So that, during a single training, we can obtain model variants for small, medium and large sizes, without additional separate training. Ablation experiments show that, compared to SISO and MIMO, LMIMO achieves a significant improvement in restoration performance with minimal computational overhead.

\begin{figure*}[t]
	\centering
	\includegraphics[width=\linewidth]{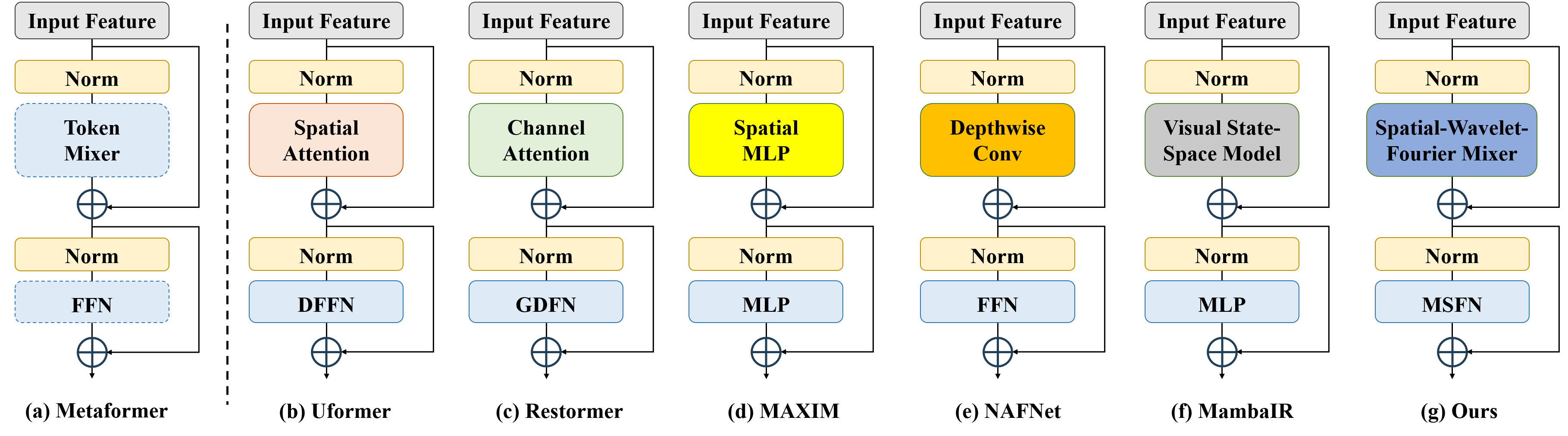}
	\caption{Comparison of Token Mixer and FFN modules across various restoration methods at Intra-block level.
	}
	\label{fig:intra-block}
\end{figure*}

\subsection{Architecture Design at Intra-block Level}
Fig.\ref{fig:intra-block} illustrates the representative intra-block level architecture design based on Transformer. In general, most current image restoration methods adopt metaformer\cite{metaformer} architecture, as shown in Fig.\ref{fig:intra-block}(a), which retains the overall structure of the Transformer while primarily modifying the Token Mixer and Feed Forward Network (FFN). As shown in Fig.\ref{fig:intra-block}, Uformer\cite{Uformer} introduces window spatial attention in the Token Mixer to reduce the high complexity of self-attention, meanwhile incorporating convolution in the FFN to provide spatial consistency. Restormer\cite{Restormer} innovatively incorporates channel self-attention in the Token Mixer to simplify computations and introduces gating mechanism in FFN. Methods like MAXIM\cite{MAXIM}, NAFNet\cite{NAFNet} and MambaIR\cite{mambair} completely discard self-attention, instead utilizing spatial MLP, depthwise convolutions and visual state-space model respectively, to achieve efficient degradation modeling.

In contrast to above approaches, we introduce multi-domain learning in Token Mixer, enabling local-region-global multi-receptive field modeling through Spatial-Wavelet-Fourier feature extraction. In FFN, we incorporate multi-scale learning to fuse Spatial-Wavelet-Fourier features at different resolutions. These fine-grained design choices allow our model to effectively balance restoration performance and model complexity, offering an efficient solution for image restoration.

\begin{figure*}[t]
	\centering
	\includegraphics[width=\linewidth]{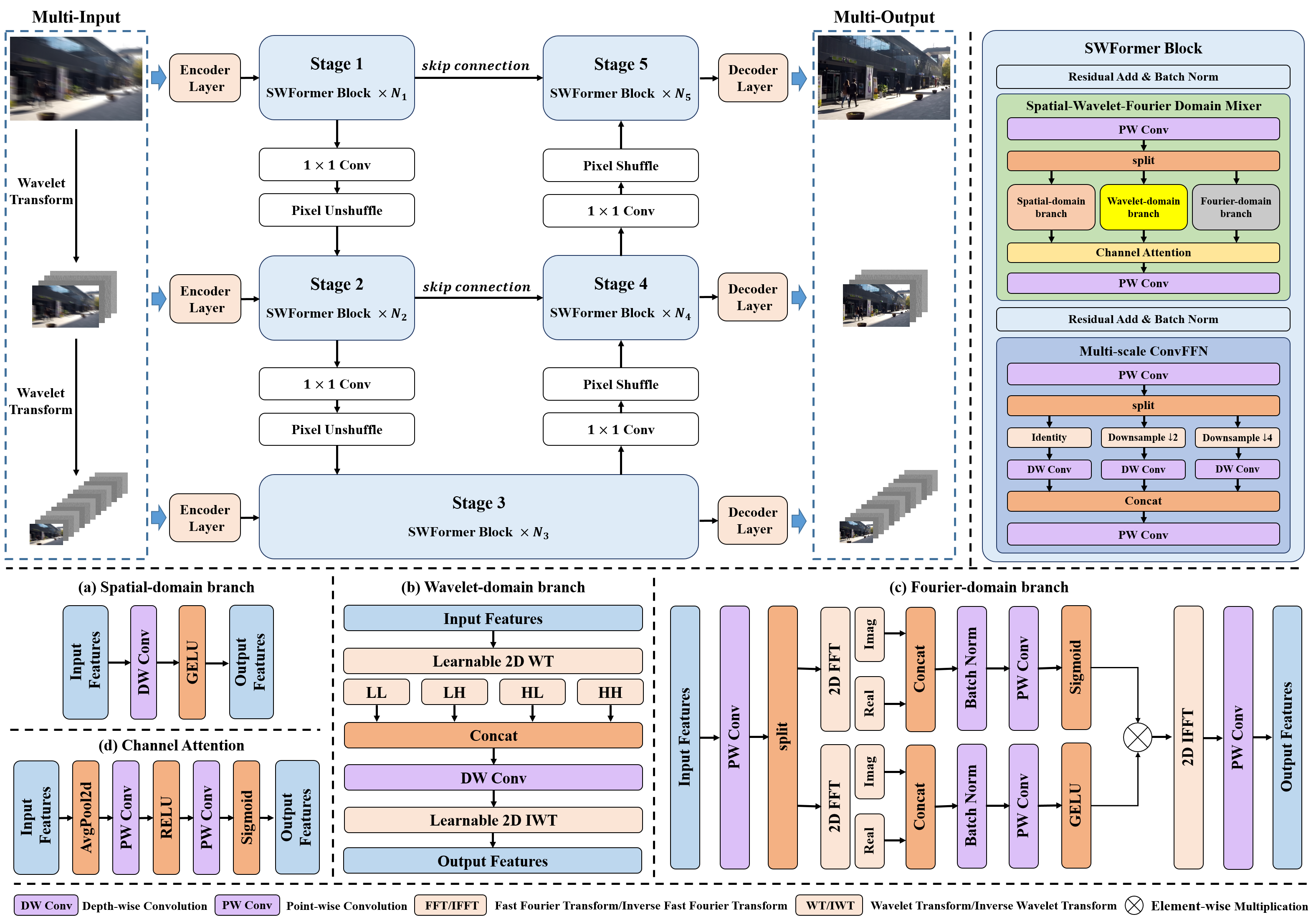}
	\caption{The overall framework of our proposed SWFormer and its detailed components.
	}
	\label{fig:method}
\end{figure*}

\section{Methods}
In this section, we first present the overall framework of our SWFormer model. We then describe the core components of the proposed SWFormer block. Finally, we provide the details of the loss functions adopted for SWFormer.

\subsection{Overall Architecture}
Building upon the previously discovered Spatial-Wavelet-Fourier priors and the inter/intra-block level architecture design, we propose the SWFormer framework, as illustrated in Fig.\ref{fig:method}. Under the LMIMO framework, SWFormer employs a hierarchical encoder-decoder structure consisting of five stages: a two-scale encoder (stage-1 and stage-2), a bottleneck (stage-3) and a two-scale decoder (stage-4 and stage-5). Given a degraded image \( I \in \mathbb{R}^{H \times W \times 3} \), SWFormer first generates multi-scale images \( \{ I^i, i=1,2,3 \} \in \{\mathbb{R}^{\frac{H}{n} \times \frac{W}{n} \times 3 \times 2^{n-1}}, n=1,2,3\} \) through Wavelet Transform, where \( H \times W \) represents the spatial dimensions and \( C \) is the number of channels. Then, \( \{ I^i, i=1,2,3 \} \) are passed through the Encoder Layer to generate input features at different scales \( \{ F^i_a, i=1,2,3 \} \in \{\mathbb{R}^{\frac{H}{n} \times \frac{W}{n} \times nC}, n=1,2,3\} \), where the Encoder Layer consists of multiple \( 3 \times 3 \) convolutions for dimensional expansion. Subsequently, \( \{ F^i_a, i=1,2,3 \} \) are passed through the five-stage hierarchical encoder-decoder structure. Between each stage, the spatial resolution is progressively reduced using Pixel Unshuffle downsampling in the two-scale encoder and then increased using Pixel Shuffle upsampling in the two-scale decoder. Within each stage, multi-scale input features \( \{ F^i_a, i=1,2,3 \} \) are processed through \( \{ N^i, i=1,2,3 \} \) SWFormer blocks to extract multi-scale output features \( \{ F^i_z, i=1, \dots, 5 \} \in \{\mathbb{R}^{\frac{H}{n} \times \frac{W}{n} \times nC}, n=1,2,3\} \). To preserve structural and textural details for restoration, low-level latent features \( \{ F^i_z, i=1,2 \} \) are concatenated with high-level latent features \( \{ F^i_z, i=4,3 \} \) through skip connections. Finally, the Decoder Layer maps \( \{ F^i_z, i=3,4,5 \} \) to the residual degraded image \( \{ R^i, i=1,2,3 \} \in \{\mathbb{R}^{\frac{H}{n} \times \frac{W}{n} \times 3 \times 2^{n-1}}, n=1,2,3\} \), which is used to obtain multi-scale clear images \( \{ O^i, i=1,2,3 \} \) via \( \{O^i = I^i + R^i, i=1,2,3 \} \), where the Decoder Layer consists of multiple \( 3 \times 3 \) convolutions for dimensional reduction. In the following two subsections, we will detail the configurations of the two core modules: SWFM and MSFN.

\subsection{Spatial-Wavelet-Fourier Mixer}
As reflected by the its name, SWFM is designed to implement a local-region-global modeling structure, consisting of three key branches: the spatial-domain branch for local perception, the wavelet-domain branch for region perception and the Fourier-domain branch for global perception. As shown in Fig.\ref{fig:method}, SWFM first applies a point-wise (PW) convolution $\widetilde{f}_{pw}$ to expand the channel dimension of the input feature \( F_0 \in \mathbb{R}^{H \times W \times C}\). Then, the expanded feature \( F_1 \in \mathbb{R}^{H \times W \times 4C} \) is split into three parts: \( F_{sp} \in \mathbb{R}^{H \times W \times C} \), \( F_{wa} \in \mathbb{R}^{H \times W \times C} \) and \( F_{fr} \in \mathbb{R}^{H \times W \times 2C} \), which are then processed through the spatial-domain branch, wavelet-domain branch and Fourier-domain branch, respectively. Next, we apply a channel-attention operation (see Fig.\ref{fig:method}(d)) to maintain feature fusion at the channel level. Finally, another point-wise convolution is applied to obtain the reduced feature \( F_2 \in \mathbb{R}^{H \times W \times C} \). The detailed implementation of the spatial-domain branch, wavelet-domain branch and Fourier-domain branch is described as follows.

\subsubsection{Spatial-domain Branch} \noindent The spatial-domain branch focuses on extracting spatial features for local modeling at the pixel level. Given the inherent local perception property of convolutions, we use depth-wise convolution to extract local spatial features. Specifically, as shown in Fig.\ref{fig:method}(a), the input feature \( F_{sp} \) in the spatial-domain branch is passed sequentially through a depth-wise convolution $\widetilde{f}_{dw}$ (DW) followed by a GELU activation function $\sigma$ to obtain the deep spatial feature \( F^d_{sp} \in \mathbb{R}^{H \times W \times C} \). The process can be expressed as:

\begin{equation}
	F^d_{sp} =  \sigma \cdot \widetilde{f}_{dw}(F_{sp})
\end{equation}

\subsubsection{Wavelet-domain Branch} \noindent The wavelet-domain branch focuses on extracting wavelet features for region-level modeling across different wavelet components. Here, we briefly revisit the wavelet transform (WT), which is widely used for analyzing the multi-scale and multi-directional characteristics of images. Given a 2D image \( x  \in \mathbb{R}^{H \times W \times 3} \), the wavelet transform \( \mathcal{W} \) (WT) decomposes it into four components $[LL, LH, HL, HH]$, which can be formulated as:

\begin{equation} \label{eq:2}
\mathcal{W}(x) = \left\{
\begin{aligned}
LL &= ( ( x \ast h ) \ast h^\top ) \downarrow_{2,2}, \\
LH &= ( ( x \ast h ) \ast g^\top ) \downarrow_{2,2}, \\
HL &= ( ( x \ast g ) \ast h^\top ) \downarrow_{2,2}, \\
HH &= ( ( x \ast g ) \ast g^\top ) \downarrow_{2,2}
\end{aligned}
\right.
\end{equation}

where \( h \) and \( g \) are low-pass and high-pass filters, \( \ast \) denotes convolution and \( \downarrow_{2,2} \) represents 2D downsampling. As shown in Eq.\ref{eq:2}, the four components \( [LL, LH, HL, HH] \) correspond to the low-frequency approximation and high-frequency details (horizontal, vertical, and diagonal) of the image, enabling effective region-level feature extraction via downsampling. The structure of the wavelet-domain branch is detailed in Fig.\ref{fig:method}(b). Specifically, the input feature \( F_{wa} \) is passed through a learnable 2D wavelet transform \( \mathcal{W}_{\phi} \) to obtain the four wavelet components \( \{ F_{LL}, F_{LH}, F_{HL}, F_{HH} \} \in \mathbb{R}^{\frac{H}{2} \times \frac{W}{2} \times C} \), where \( \phi \) represents parameterized convolutions. These components are concatenated along the channel dimension and processed through a depth-wise convolution \( \widetilde{f}_{dw} \) for feature extraction in the wavelet domain. Finally, a learnable 2D inverse wavelet transform (IWT) reconstructs the modulated deep wavelet feature \( F^d_{wa} \in \mathbb{R}^{H \times W \times C} \). This process can be formulated as:

\begin{equation} \label{eq:3}
F^d_{wa} = \mathcal{W}^{-1}\left( \sigma \cdot \widetilde{f}_{dw} \left( \text{Concat}[F_{LL}, F_{LH}, F_{HL}, F_{HH}] \right) \right)
\end{equation}

\subsubsection{Fourier-domain Branch} \noindent The frequency-domain branch focuses on capturing frequency features for global modeling around the entire image. Here, we review the Fourier transform, which is widely used for analyzing the frequency characteristic of images. Given an image $x$, the fast Fourier transform $\mathscr{F}$ converts it to frequency space as the complex component $\mathscr{F}(x)$, which is expressed as:

\begin{equation} \label{eq:fft}
	\mathscr{F}(x)(u,v) = \frac{1}{\sqrt{HW}} \sum\limits_{h=0}^{H-1} \sum\limits_{w=0}^{W-1} x(h, w)e^{-j2\pi(\frac{h}{H}u+\frac{w}{W}v)}
\end{equation}

where $u$ and $v$ are the coordinates of the Fourier space. $\mathscr{F}^{-1}$ represents the inverse fast Fourier transform (IFFT). As shown in Eq.\ref{eq:fft}, individual components in the frequency domain correspond to entire pixel sets in the spatial domain. Unlike the frequency dynamic convolutions in the conference version, which improve restoration performance but slow down inference speed, we adopt a gating mechanism to filter the frequency components directly. As illustrated in Fig.\ref{fig:method}(c), the input feature \( F_{fr} \) is passed through a PW convolution and split into two gating vectors: \( F_{fr}^1 \in \mathbb{R}^{H \times W \times C} \) and \( F_{fr}^2 \in \mathbb{R}^{H \times W \times C} \). We then apply 2D FFT to map the features to the Fourier domain and concatenate the real and imaginary parts, producing joint features \( F_j^1 \in \mathbb{R}^{H \times W \times 2C} \) and \( F_j^2 \in \mathbb{R}^{H \times W \times 2C} \). \( F_j^1 \) undergoes batch normalization, a PW convolution and a GELU activation function to obtain frequency features \( F_{fd} \in \mathbb{R}^{H \times W \times 2C} \), while \( F_j^2 \) is processed with batch normalization, a PW convolution and a Sigmoid activation to produce gating features \( F_{ga} \in \mathbb{R}^{H \times W \times 2C} \). Finally, the two features are element-wise multiplied, mapped back to the spatial domain via 2D IFFT, and reduced using a PW convolution $\widetilde{f}_{pw}$ to obtain the modulated deep Fourier feature \( F_{fr}^d  \in \mathbb{R}^{H \times W \times C}\). This process can be formulated as:

\begin{equation} \label{eq:fr}
F_{fr}^d = \widetilde{f}_{pw} \left(\mathscr{F}^{-1} \left( F_j^1 \odot F_j^2 \right) \right)
\end{equation}

\subsection{Multi-Scale ConvFFN}
The Feed-Forward Network is a crucial module in Transformers, and in this work, we use it to integrate the local-region-global features extracted by the token mixer. Specifically, we introduce multi-scale learning to fuse features across different resolutions. As shown in Fig.\ref{fig:method}, MSFN starts by applying a PW convolution to double the dimension of the input features \( F_2 \). The expanded features \( F_3 \in \mathbb{R}^{H \times W \times 2C} \) are split into three parts: one part undergoes identity operation, another is downsampled by a factor of two, and the third is downsampled by a factor of four. These components are processed through DW convolutions to capture local connections. After aggregating the multi-scale features, a final PW convolution reduces the dimension. This design efficiently fuses features from multiple scales, capturing rich spatial information while maintaining computational efficiency.

\subsection{Loss Function}
To align with the model design, we introduce a multi-domain loss in the optimization flow, as shown in Eq.\ref{eq:loss}. This loss consists of three components: spatial-domain, wavelet-domain and Fourier-domain losses.

\begin{align} \label{eq:loss}
    L^i &=  \left \| O^i - G^i  \right \|_{1} 
        + \left \| \mathcal{W}(O^i) - \mathcal{W}(G^i)  \right \|_{1} \notag \\
        &\quad + \lambda\left \| \mathscr{F}(O^i) - \mathscr{F}(G^i)  \right \|_{1}, 
        \quad i=1,2,3
\end{align}

Here, the L1 loss regularizes \( \{O^i, i=1,2,3\} \) to match multi-scale ground truth \( \{G^i, i=1,2,3\} \). The hyperparameter \( \lambda \) is set to 0.1 to balance the contributions of each domain loss.

\section{Experimental Results}
To evaluate our SWFormer effectiveness, we conduct extensive experiments on common image restoration tasks. In tables, the best and second-best quality scores of the evaluated methods are \textbf{highlighted} and \underline{underlined}. 

\subsection{Experimental Settings}

\subsubsection{Datasets} \noindent We adopt RESIDE\cite{reside}, O-HAZE\cite{Ohaze}, NH-HAZE\cite{nhHAZE} and DENSE-HAZE\cite{denseHAZE} for dehazing; Rain200H\cite{RAIN200}, Rain200L\cite{RAIN200}, DDN-Data\cite{ddn}, DID-Data\cite{DID} and SPA-Data\cite{SPA-Data} for deraining; Raindrop\cite{AttentiveGANraindrop} for raindrop removal; CSD\cite{HDCW-NetCSD}, SRRS\cite{JSTASRSRRS} and Snow100K\cite{desnownesnow100k} for desnowing; UIEB\cite{waternetuieb} and LSUI\cite{U-shapeTranslsui} for underwater enhancement; LOL-v1\cite{lolv1} and LOL-v2\cite{sparselolv2} for low-light enhancement; GoPro\cite{GOPRO} and HIDE\cite{HIDE} for motion deblurring; DPDD\cite{DPDD} and LFDOF\cite{aifnet} for defocus deblurring; AISTD\cite{aistd} for shadow removal; CUHK-CR\cite{de-msda_memory} for cloud removal. More details of dataset scales and resolutions can be found in Supplementary material.

\subsubsection{Implementation Details} \noindent  AdamW optimizer\cite{adamw} with $\beta_1$ and $\beta_2$ equal to 0.9 and 0.999 is used to train SWFormer. The initial learning rate is set as $10^{-3}$. We adopt the cosine annealing strategy\cite{cos} to train the models, where the learning rate gradually decreases from the initial learning rate to $10^{-6}$. All experiments are implemented by PyTorch\cite{pytorch} 1.7.1 with four NVIDIA 3090 GPUs. More implementation details for each restoration tasks can be found in Supplementary material.

\subsection{Results on Rain Streak/Raindrop Removal Task}
\begin{table}[!b]
\setlength{\abovecaptionskip}{0cm}
\setlength{\belowcaptionskip}{-0.3cm}
\renewcommand{\arraystretch}{1}
\centering
\caption{Quantitative evaluations on raindrop removal.}\label{tab:raindrop}
\resizebox{0.45\textwidth}{!}{
\begin{tabular}{clllllll|clcl|clcl}
\hline
\multicolumn{8}{c|}{\multirow{2}{*}{Methods}} & \multicolumn{4}{c|}{Raindrop-A\cite{AttentiveGANraindrop}}                        & \multicolumn{4}{c}{Raindrop-B\cite{AttentiveGANraindrop}}                                                \\ \cline{9-16} 
\multicolumn{8}{c|}{}                         & \multicolumn{2}{c}{PSNR$\uparrow$}  & \multicolumn{2}{c|}{SSIM$\uparrow$}  & \multicolumn{2}{c}{PSNR$\uparrow$}  & \multicolumn{2}{c}{SSIM$\uparrow$} \\ \hline
% \multicolumn{8}{c|}{(CVPR'17)pix2pix\cite{pix2pix}}                  & \multicolumn{2}{c}{28.02} & \multicolumn{2}{c|}{0.855} & \multicolumn{2}{c}{-}     & \multicolumn{2}{c}{-}        \\
\multicolumn{8}{c|}{(CVPR'19)DuRN\cite{DuRN}}                     & \multicolumn{2}{c}{31.24} & \multicolumn{2}{c|}{0.926} & \multicolumn{2}{c}{25.32} & \multicolumn{2}{c}{0.817}     \\
% \multicolumn{8}{c|}{(ICCV'19)RaindropAttn\cite{RaindropAttn}}             & \multicolumn{2}{c}{31.44} & \multicolumn{2}{c|}{0.926} & \multicolumn{2}{c}{-}     & \multicolumn{2}{c}{-}          \\
\multicolumn{8}{c|}{(CVPR'18)AttentiveGAN\cite{AttentiveGANraindrop}}             & \multicolumn{2}{c}{31.59} & \multicolumn{2}{c|}{0.917} & \multicolumn{2}{c}{25.05} & \multicolumn{2}{c}{0.811}       \\
\multicolumn{8}{c|}{(TPAMI'22)IDT\cite{IDT}}                      & \multicolumn{2}{c}{31.87} & \multicolumn{2}{c|}{0.931} & \multicolumn{2}{c}{-}     & \multicolumn{2}{c}{-}         \\
\multicolumn{8}{c|}{(CVPR'22)MAXIM\cite{MAXIM}}                 & \multicolumn{2}{c}{31.87} & \multicolumn{2}{c|}{0.935} & \multicolumn{2}{c}{25.74} & \multicolumn{2}{c}{0.827}     \\
\multicolumn{8}{c|}{(TPAMI'23)RainDropDiff128\cite{RainDropDiff128}}          & \multicolumn{2}{c}{32.43} & \multicolumn{2}{c|}{0.933} & \multicolumn{2}{c}{-}     & \multicolumn{2}{c}{-}         \\ 
\multicolumn{8}{c|}{(ICCV'23)UDR-$\mathrm{S^2}$Former\cite{udrsformer}}                         & \multicolumn{2}{c}{32.64}      & \multicolumn{2}{c|}{0.943}      & \multicolumn{2}{c}{26.92}      & \multicolumn{2}{c}{0.832}          \\ 
\multicolumn{8}{c|}{(CVPR'24)AST\cite{ast}}                         & \multicolumn{2}{c}{32.45}      & \multicolumn{2}{c|}{0.937}      & \multicolumn{2}{c}{24.99}      & \multicolumn{2}{c}{0.806}          \\
\multicolumn{8}{c|}{(ECCV'24)SFHformer\cite{sfhformer}}                         & \multicolumn{2}{c}{33.10}      & \multicolumn{2}{c|}{\underline{0.946}}      & \multicolumn{2}{c}{27.17}      & \multicolumn{2}{c}{\textbf{0.838}}          \\ \hline
\multicolumn{8}{c|}{(Ours)SWFormer-s}                         & \multicolumn{2}{c}{33.29}      & \multicolumn{2}{c|}{0.945}      & \multicolumn{2}{c}{27.34}      & \multicolumn{2}{c}{0.834}          \\ 
\multicolumn{8}{c|}{(Ours)SWFormer-m}                         & \multicolumn{2}{c}{\underline{33.50}}      & \multicolumn{2}{c|}{\underline{0.946}}      & \multicolumn{2}{c}{\underline{27.44}}      & \multicolumn{2}{c}{\underline{0.835}}          \\ 
\multicolumn{8}{c|}{(Ours)SWFormer-l}                         & \multicolumn{2}{c}{\textbf{33.55}}      & \multicolumn{2}{c|}{\textbf{0.948}}      & \multicolumn{2}{c}{\textbf{27.48}}      & \multicolumn{2}{c}{\underline{0.835}}          \\ \hline
\end{tabular}}
\end{table}

\begin{table*}[!t]
\setlength{\abovecaptionskip}{0.0cm}
\setlength{\belowcaptionskip}{0.0cm}
\renewcommand{\arraystretch}{1}
\centering
\begin{minipage}[c]{\textwidth}
\captionof{table}{Quantitative evaluations on synthetic and real-world deraining datasets.}\label{tab:rain}
\resizebox{\textwidth}{!}{
\setlength{\tabcolsep}{1.1mm}{
\begin{tabular}{cllll|ccl|clcl|clcl|clcl|clcl}
\hline
\multicolumn{5}{c|}{\multirow{3}{*}{Method}}      & \multicolumn{15}{c|}{Synthetic}  & \multicolumn{4}{c}{Real}  \\ \cline{6-24}
\multicolumn{5}{c|}{}        & \multicolumn{3}{c|}{Rain200L\cite{RAIN200}}                                & \multicolumn{4}{c|}{Rain200H\cite{RAIN200}}                                & \multicolumn{4}{c|}{DDN-Data\cite{ddn}}           & \multicolumn{4}{c|}{DID-Data\cite{DID}}      & \multicolumn{4}{c}{SPA-Data\cite{SPA-Data}}          \\ \cline{6-24}
\multicolumn{5}{c|}{}              & \multicolumn{1}{c}{PSNR$\uparrow$}  & \multicolumn{2}{c|}{SSIM$\uparrow$}   & \multicolumn{2}{c}{PSNR$\uparrow$}  & \multicolumn{2}{c|}{SSIM$\uparrow$}   & \multicolumn{2}{c}{PSNR$\uparrow$}  & \multicolumn{2}{c|}{SSIM$\uparrow$}   & \multicolumn{2}{c}{PSNR$\uparrow$}  & \multicolumn{2}{c|}{SSIM$\uparrow$} & \multicolumn{2}{c}{PSNR$\uparrow$}  & \multicolumn{2}{c}{SSIM$\uparrow$}    \\ \hline
\multicolumn{5}{c|}{(CVPR'21)MPRNet\cite{MPRNet}}    & \multicolumn{1}{c}{39.47} & \multicolumn{2}{c|}{0.9825}       & \multicolumn{2}{c}{30.67} & \multicolumn{2}{c|}{0.9110} & \multicolumn{2}{c}{33.10} & \multicolumn{2}{c|}{0.9347} & \multicolumn{2}{c}{33.99} & \multicolumn{2}{c|}{0.9590} & \multicolumn{2}{c}{43.64} & \multicolumn{2}{c}{0.9844} \\
\multicolumn{5}{c|}{(AAAI'21)DualGCN\cite{dualgcn} }   & \multicolumn{1}{c}{40.73} & \multicolumn{2}{c|}{0.9886}     & \multicolumn{2}{c}{31.15} & \multicolumn{2}{c|}{0.9125} & \multicolumn{2}{c}{33.01} & \multicolumn{2}{c|}{0.9489} & \multicolumn{2}{c}{34.37} & \multicolumn{2}{c|}{0.9620} & \multicolumn{2}{c}{44.18} & \multicolumn{2}{c}{0.9902}  \\
\multicolumn{5}{c|}{(ICCV'21)SPDNet\cite{SPDNet}}   & \multicolumn{1}{c}{40.50} & \multicolumn{2}{c|}{0.9875}     & \multicolumn{2}{c}{31.28} & \multicolumn{2}{c|}{0.9207} & \multicolumn{2}{c}{33.15}     & \multicolumn{2}{c|}{0.9457}      & \multicolumn{2}{c}{34.57}      & \multicolumn{2}{c|}{0.9560}       & \multicolumn{2}{c}{43.20}      & \multicolumn{2}{c}{0.9871}      \\
\multicolumn{5}{c|}{(CVPR'22)Uformer\cite{Uformer}}    & \multicolumn{1}{c}{40.20} & \multicolumn{2}{c|}{0.9860}     & \multicolumn{2}{c}{30.80} & \multicolumn{2}{c|}{0.9105} & \multicolumn{2}{c}{33.95} & \multicolumn{2}{c|}{0.9545} & \multicolumn{2}{c}{35.02}      & \multicolumn{2}{c|}{0.9621}       & \multicolumn{2}{c}{46.13}      & \multicolumn{2}{c}{0.9913}           \\
\multicolumn{5}{c|}{(CVPR'22)Restormer\cite{Restormer}}   & \multicolumn{1}{c}{40.99} & \multicolumn{2}{c|}{0.9890}    & \multicolumn{2}{c}{32.00} & \multicolumn{2}{c|}{0.9329} & \multicolumn{2}{c}{34.20} & \multicolumn{2}{c|}{0.9571}  & \multicolumn{2}{c}{35.29}     & \multicolumn{2}{c|}{0.9641}      & \multicolumn{2}{c}{47.98}     & \multicolumn{2}{c}{0.9921}     \\ 
\multicolumn{5}{c|}{(TPAMI'22)IDT\cite{IDT}}    & \multicolumn{1}{c}{40.74} & \multicolumn{2}{c|}{0.9884}    & \multicolumn{2}{c}{32.10}      & \multicolumn{2}{c|}{0.9344}       & \multicolumn{2}{c}{33.84}      & \multicolumn{2}{c|}{0.9549}       & \multicolumn{2}{c}{34.89}      & \multicolumn{2}{c|}{0.9623}       & \multicolumn{2}{c}{47.35}      & \multicolumn{2}{c}{0.9930}          \\
\multicolumn{5}{c|}{(AAAI'23)HCT-FFN\cite{HCTFFN}}   & \multicolumn{1}{c}{39.70} & \multicolumn{2}{c|}{0.9850}     & \multicolumn{2}{c}{31.51}      & \multicolumn{2}{c|}{0.9100}       & \multicolumn{2}{c}{33.00}      & \multicolumn{2}{c|}{0.9502}       & \multicolumn{2}{c}{33.96}      & \multicolumn{2}{c|}{0.9592}       & \multicolumn{2}{c}{45.79}      & \multicolumn{2}{c}{0.9898}        \\
\multicolumn{5}{c|}{(CVPR'23)DRSformer\cite{DRSformer}}    & \multicolumn{1}{c}{41.23} & \multicolumn{2}{c|}{0.9894}    & \multicolumn{2}{c}{32.17}      & \multicolumn{2}{c|}{0.9326}       & \multicolumn{2}{c}{34.35}      & \multicolumn{2}{c|}{0.9588}       & \multicolumn{2}{c}{35.35}      & \multicolumn{2}{c|}{0.9646}       & \multicolumn{2}{c}{48.54}      & \multicolumn{2}{c}{0.9924}         \\
\multicolumn{5}{c|}{(CVPR'24)NeRD-Rain\cite{nerd-rain}}    & \multicolumn{1}{c}{41.71} & \multicolumn{2}{c|}{0.9903}    & \multicolumn{2}{c}{32.40}      & \multicolumn{2}{c|}{0.9373}       & \multicolumn{2}{c}{\underline{34.45}}      & \multicolumn{2}{c|}{0.9596}       & \multicolumn{2}{c}{\underline{35.53}}      & \multicolumn{2}{c|}{\underline{0.9659}}       & \multicolumn{2}{c}{49.58}      & \multicolumn{2}{c}{0.9940}             \\
\multicolumn{5}{c|}{(ECCV'24)FADformer\cite{fadformer}}  & \multicolumn{1}{c}{41.80} & \multicolumn{2}{c|}{0.9906} & \multicolumn{2}{c}{32.48}      & \multicolumn{2}{c|}{0.9359}      & \multicolumn{2}{c}{34.42}      & \multicolumn{2}{c|}{\underline{0.9602}}       & \multicolumn{2}{c}{35.48}      & \multicolumn{2}{c|}{0.9657}       & \multicolumn{2}{c}{49.21}      & \multicolumn{2}{c}{0.9934}          \\
\multicolumn{5}{c|}{(ECCV'24)SFHformer\cite{sfhformer}}  & \multicolumn{1}{c}{41.85} & \multicolumn{2}{c|}{0.9908} & \multicolumn{2}{c}{32.33}      & \multicolumn{2}{c|}{0.9351}      & \multicolumn{2}{c}{34.38}      & \multicolumn{2}{c|}{0.9594}       & \multicolumn{2}{c}{35.44}      & \multicolumn{2}{c|}{0.9655}       & \multicolumn{2}{c}{\underline{50.11}}      & \multicolumn{2}{c}{\underline{0.9942}}        \\ \hline
\multicolumn{5}{c|}{(Ours)SWFormer-s}    & \multicolumn{1}{c}{42.01} & \multicolumn{2}{c|}{0.9908}    & \multicolumn{2}{c}{32.43}      & \multicolumn{2}{c|}{0.9345}       & \multicolumn{2}{c}{34.18}      & \multicolumn{2}{c|}{0.9573}       & \multicolumn{2}{c}{35.18}      & \multicolumn{2}{c|}{0.9639}       & \multicolumn{2}{c}{49.03}      & \multicolumn{2}{c}{0.9936}           \\
\multicolumn{5}{c|}{(Ours)SWFormer-m}    & \multicolumn{1}{c}{\underline{42.23}} & \multicolumn{2}{c|}{\underline{0.9913}}    & \multicolumn{2}{c}{\underline{32.69}}      & \multicolumn{2}{c|}{\underline{0.9377}}       & \multicolumn{2}{c}{34.36}      & \multicolumn{2}{c|}{0.9592}       & \multicolumn{2}{c}{35.42}      & \multicolumn{2}{c|}{0.9654}       & \multicolumn{2}{c}{49.66}      & \multicolumn{2}{c}{0.9941}            \\
\multicolumn{5}{c|}{\textbf{(Ours)SWFormer-l}}  & \multicolumn{1}{c}{\textbf{42.70}} & \multicolumn{2}{c|}{\textbf{0.9922}} & \multicolumn{2}{c}{\textbf{33.09}}      & \multicolumn{2}{c|}{\textbf{0.9441}}       & \multicolumn{2}{c}{\textbf{34.51}}      & \multicolumn{2}{c|}{\textbf{0.9608}}       & \multicolumn{2}{c}{\textbf{35.58}}      & \multicolumn{2}{c|}{\textbf{0.9664}}       & \multicolumn{2}{c}{\textbf{50.28}}      & \multicolumn{2}{c}{\textbf{0.9943}}            \\ \hline
\end{tabular}}
}
\vspace{1mm}
\end{minipage}

\begin{minipage}[c]{\textwidth}
\includegraphics[width=\linewidth]{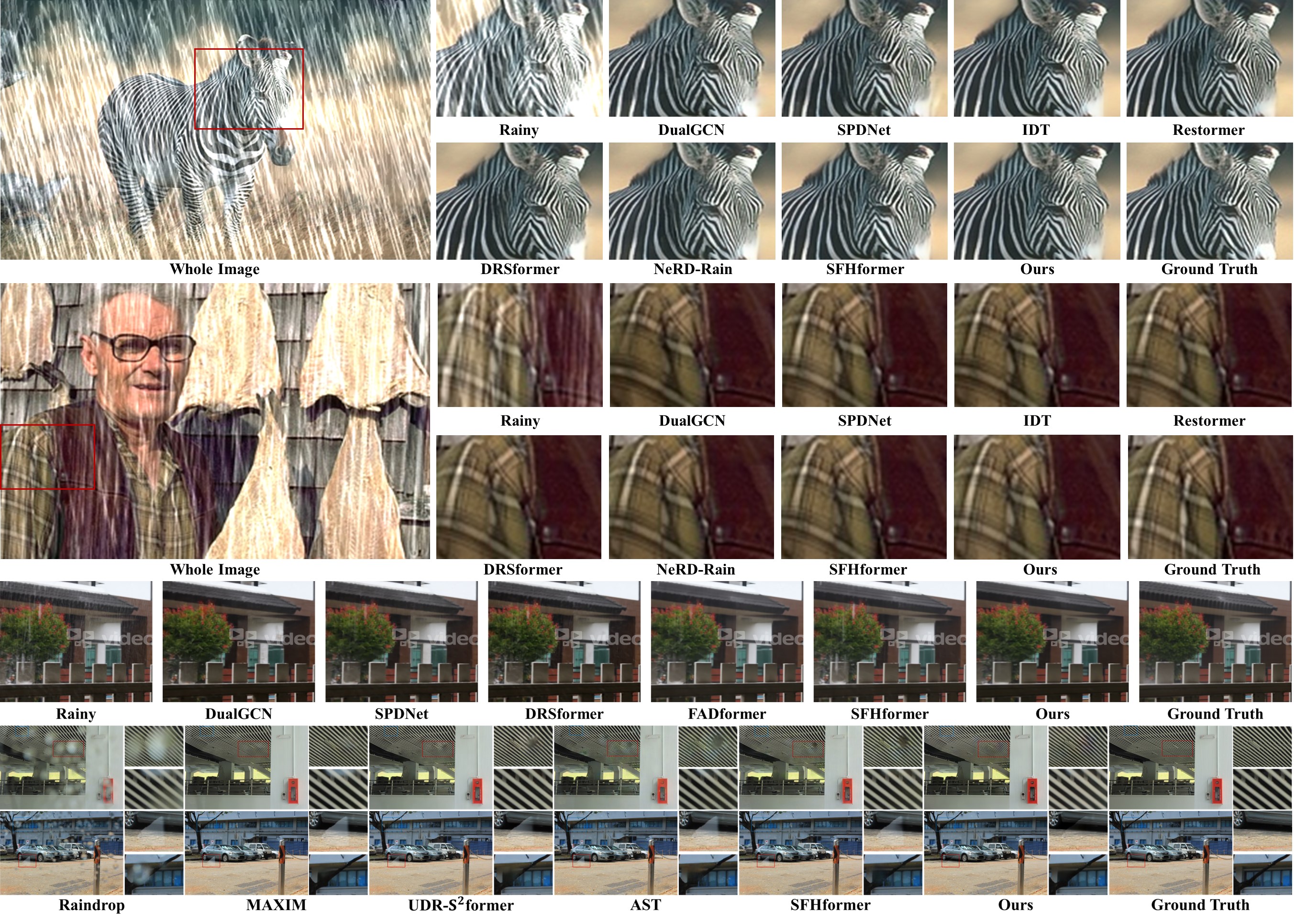}
\captionof{figure}{The quantitative evaluation results on synthetic/real-world rain streak removal and raindrop removal.} \label{fig:rain}
\vspace{-3mm}
\end{minipage}
\end{table*}

We conduct rain streak/raindrop removal experiments on six widely used public datasets, including four synthetic datasets, Rain200L\cite{RAIN200}, Rain200H\cite{RAIN200}, DID-Data\cite{DID} and DDN-Data\cite{ddn}, and two real-world datasets, SPA-Data \cite{SPA-Data} and Raindrop \cite{AttentiveGANraindrop}. Tab.\ref{tab:rain} presents the quantitative comparison results for rain streak removal. Compared to the latest SOTA methods, our proposed SWFormer achieves the best performance in terms of PSNR and SSIM across all datasets. Notably, on the Rain200L and Rain200H datasets, even our smallest model, SWFormer-small, outperforms SFHFormer\cite{sfhformer} in PSNR. Tab.\ref{tab:raindrop} also presents the quantitative results for raindrop removal, where SWFormer demonstrates a highly competitive performance advantage. Fig.\ref{fig:rain} shows the qualitative comparison results for synthetic rain streak removal, real-world rain streak removal and raindrop removal, presented from top to bottom. Our method consistently delivers superior rain removal results, excelling in texture detail preservation and raindrop repair. In contrast, other methods either fail to effectively remove rain or over-remove it, mistakenly eliminating white stripes or clothing bands as rain streaks.

\subsection{Results on Low-light Image Enhancement Task}

\begin{table*}[!t]
\setlength{\abovecaptionskip}{0cm}
\setlength{\belowcaptionskip}{0cm}
\renewcommand{\arraystretch}{1.1}
\centering
\begin{minipage}[c]{\textwidth}
\captionof{table}{Quantitative evaluations on synthetic and real-world low-light enhancement.\label{tab:low-light}}
\resizebox{\textwidth}{!}{
\setlength{\tabcolsep}{1mm}{
\begin{tabular}{c|cccc|cccc|cccc}
\hline
\multirow{3}{*}{Method} & \multicolumn{4}{c|}{LOL-v1\cite{lolv1}}                               & \multicolumn{4}{c|}{LOL-v2-real\cite{sparselolv2}}                          & \multicolumn{4}{c}{LOL-v2-syn\cite{sparselolv2}}                           \\
                        & \multicolumn{2}{c}{Normal} & \multicolumn{2}{c|}{GT Mean} & \multicolumn{2}{c}{Normal} & \multicolumn{2}{c|}{GT Mean} & \multicolumn{2}{c}{Normal} & \multicolumn{2}{c}{GT Mean} \\ \cline{2-13} 
                        & PSNR$\uparrow$         & SSIM$\uparrow$        & PSNR$\uparrow$          & SSIM$\uparrow$         & PSNR$\uparrow$         & SSIM$\uparrow$        & PSNR$\uparrow$          & SSIM$\uparrow$         & PSNR$\uparrow$         & SSIM$\uparrow$        & PSNR$\uparrow$        & SSIM$\uparrow$         \\ \hline
% (ACMMM'19)KinD\cite{kindacmmm2019}                    & 17.65        & 0.775       & 20.86         & 0.802        & 14.74        & 0.641       & 17.54         & 0.669        & 13.29        & 0.578       & 16.26        & 0.591        \\
% (CVPR'20)ZeroDCE\cite{zerodcecvpr2020}                 & 14.86        & 0.559       & 21.88         & 0.640        & 16.06        & 0.580       & 19.77         & 0.671        & 17.71        & 0.815       & 21.46        & 0.848        \\
(TPAMI'20)3DLUT\cite{3DLUTtpami2020}                   & 14.35        & 0.445       & 21.35         & 0.585        & 17.59        & 0.721       & 20.19         & 0.745        & 18.04        & 0.800       & 22.17        & 0.854        \\
(TIP'21)Sparse\cite{sparselolv2}                  & 17.20        & 0.640       & -             & -            & 20.06        & 0.816       & -             & -            & 22.05        & 0.905       & -            & -            \\
% (TIP'21)DRBN\cite{DRBN}                    & 16.29        & 0.617       & 19.55         & 0.746        & 20.29        & 0.831       & -             & -            & 23.22        & 0.927       & -            & -            \\
(CVPR'21)RUAS\cite{RUAS}                    & 16.41        & 0.500       & 18.65         & 0.518        & 15.33        & 0.488       & 19.06         & 0.510        & 13.77        & 0.638       & 16.58        & 0.719        \\
(AAAI'22)LLFlow\cite{llflowaaai22}                  & 21.15        & 0.854       & 24.99         & 0.871        & 17.43        & 0.831       & 25.42         & 0.877        & 24.81        & 0.919       & 27.96        & 0.930        \\
% (TIP'21)EnGAN\cite{engan}                   & 17.48        & 0.651       & 20.00         & 0.691        & 18.23        & 0.617       & -             & -            & 16.57        & 0.734       & -            & -            \\
(CVPR'22)Restormer\cite{Restormer}               & 22.37        & 0.816       & 26.68         & 0.853        & 18.69        & 0.834       & 26.12         & 0.853        & 21.41        & 0.830       & 25.43        & 0.859        \\
(ECCV'22)LEDNet\cite{ledneteccv2022}                  & 20.63        & 0.823       & 25.47         & 0.846        & 19.94        & 0.827       & 27.81         & 0.870        & 23.71        & 0.914       & 27.37        & 0.928        \\
(CVPR'22)SNR-Net\cite{snrnet}                 & 24.61        & 0.842       & 26.72         & 0.851        & 21.48        & 0.849       & 27.21         & 0.871        & 24.14        & 0.928       & 27.79        & 0.941        \\
(CVPR'23)PairLIE\cite{pairliecvpr2023}                 & 19.51        & 0.736       & 23.53         & 0.755        & 19.89        & 0.778       & 24.03         & 0.803        & -            & -           & -            & -            \\
(AAAI'23)LLFormer\cite{llformeraaai2023}                & 23.65        & 0.816       & 25.76         & 0.823        & 20.06        & 0.792       & 26.20         & 0.819        & 24.04        & 0.909       & 28.01        & 0.927        \\
(ICCV'23)RetinexFormer\cite{retinexformer}           & 25.15        & 0.846       & 27.14         & 0.850        & 22.79        & 0.840       & 27.69         & 0.856        & 25.67        & 0.930       & 28.99        & 0.939        \\
(ECCV'24)SFHformer\cite{sfhformer}               & 24.29        & 0.862       &       26.97       &    0.869          & \underline{23.78}        & \underline{0.872}       &      \underline{28.36}         &       \underline{0.883}       & 25.80        & 0.937       &     29.06         &      0.940        \\
(CVPR'25)CIDNet\cite{cidnetcvpr2025}                  & 23.50        & \underline{0.870}       & \textbf{28.14}         & \underline{0.889}        & 23.43        & 0.862       & 27.76         & 0.881        & 25.71        & \underline{0.942}       & 29.57        & \underline{0.950}        \\ \hline
(Ours)SWFormer-s              &       25.26       &       0.861      &         27.52      &       0.864       & 22.79        & 0.854       & 28.51         & 0.879        & 26.17        & 0.932       & 29.31        & 0.941        \\
(Ours)SWFormer-m              &        \underline{25.65}      &       \underline{0.870}      &       \underline{27.96}        &       0.882       & 23.42        & 0.854       & 28.32         & 0.876        & \underline{26.55}        & 0.934       & \underline{29.63}        & 0.944        \\
(Ours)SWFormer-l              &     \textbf{25.70}         &     \textbf{0.876}        &       \textbf{28.14}        &        \textbf{0.892}      & \textbf{23.86}        & \textbf{0.877}       & \textbf{28.72}         & \textbf{0.896}        & \textbf{27.01}        & \textbf{0.943}       & \textbf{30.26}        & \textbf{0.952}        \\ \hline
\end{tabular}}}
\vspace{1mm}
\end{minipage}
\begin{minipage}[c]{\textwidth}
\includegraphics[width=\linewidth]{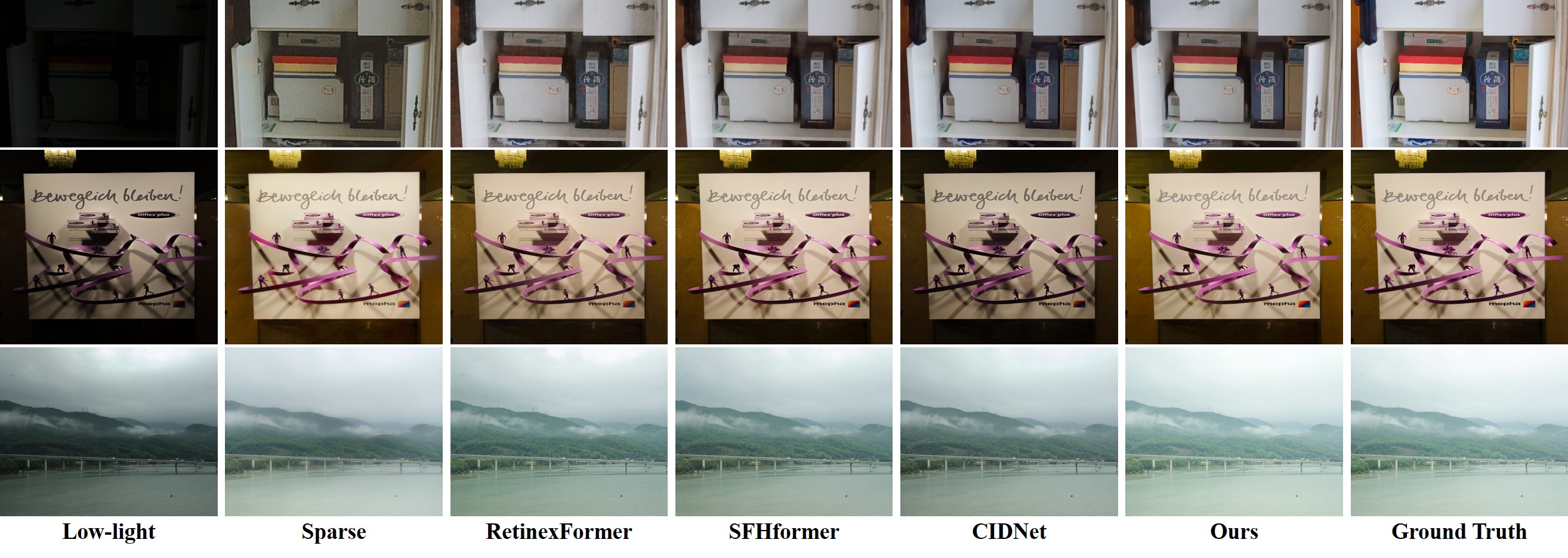}
\captionof{figure}{The quantitative evaluation results on  synthetic and real-world low-light enhancement.} \label{fig:lowlight}
\vspace{-2mm}
\end{minipage}
\end{table*}

We conduct low-light image enhancement experiments on three widely recognized public datasets: LOLv1\cite{lolv1}, LOLv2-real\cite{sparselolv2} (real-world low-light dataset) and LOLv2-syn\cite{sparselolv2} (synthetic low-light dataset). Tab.\ref{tab:low-light} presents the quantitative comparison results across these datasets. To provide a comprehensive evaluation of our approach, we compute PSNR and SSIM using two commonly employed methods: the standard computation and a variant with GT Mean preprocessing\cite{cidnetcvpr2025}. Compared to the most recent state-of-the-art methods, our proposed SWFormer consistently achieves superior performance on all three datasets, using both calculation methods. Specifically, our model outperforms SFHFormer\cite{sfhformer}, demonstrating PSNR improvements of 1.41 dB and 1.21 dB on the LOLv1 and LOLv2-syn datasets, respectively. Fig.\ref{fig:lowlight} displays the qualitative comparison results for low-light enhancement. Our method excels in light correction and noise removal, producing the most visually appealing and realistic results. In contrast, RetinexFormer\cite{retinexformer} struggles with noise removal, failing to maintain a clean image, while CIDNet\cite{cidnetcvpr2025} demonstrates limitations in light correction, leaving the restored images with unnatural lighting.

\subsection{Results on Image Dehazing Task}
\begin{table*}[!t]
\setlength{\abovecaptionskip}{0cm}
\setlength{\belowcaptionskip}{0cm}
\renewcommand{\arraystretch}{1}
\centering
\begin{minipage}[c]{\textwidth}
\captionof{table}{Quantitative evaluations on the synthetic and real-world dehazing.\label{tab:haze}}
\resizebox{\textwidth}{!}{
\setlength{\tabcolsep}{0.95mm}{
\begin{tabular}{clllllll|clcl|clcl|clcl|clcl|clcl}
\hline
\multicolumn{8}{c|}{\multirow{3}{*}{Method}} & \multicolumn{4}{c|}{ITS\cite{reside}}                                & \multicolumn{4}{c|}{OTS\cite{reside}}                                & \multicolumn{4}{c|}{\multirow{2}{*}{\makecell{O-HAZE\cite{Ohaze}}}}          & \multicolumn{4}{c|}{\multirow{2}{*}{\makecell{NH-HAZE\cite{nhHAZE}}}}         & \multicolumn{4}{c}{\multirow{2}{*}{\makecell{DENSE-HAZE\cite{denseHAZE}}}}               \\ \cline{9-16}
\multicolumn{8}{c|}{}                        & \multicolumn{4}{c|}{SOTS-indoor}                        & \multicolumn{4}{c|}{SOTS-outdoor}                       & \multicolumn{4}{c|}{}                                 & \multicolumn{4}{c|}{}                                 & \multicolumn{4}{c}{}                                                                    \\ \cline{9-28} 
\multicolumn{8}{c|}{}                        & \multicolumn{2}{c}{PSNR$\uparrow$}  & \multicolumn{2}{c|}{SSIM$\uparrow$}   & \multicolumn{2}{c}{PSNR$\uparrow$}  & \multicolumn{2}{c|}{SSIM$\uparrow$}   & \multicolumn{2}{c}{PSNR$\uparrow$}  & \multicolumn{2}{c|}{SSIM$\uparrow$} & \multicolumn{2}{c}{PSNR$\uparrow$}  & \multicolumn{2}{c|}{SSIM$\uparrow$} & \multicolumn{2}{c}{PSNR$\uparrow$}  & \multicolumn{2}{c}{SSIM$\uparrow$}  \\ \hline
% \multicolumn{8}{c|}{(TIP'16)DehazeNet\cite{DehazeNet}}       & \multicolumn{2}{c}{19.82} & \multicolumn{2}{c|}{0.8210} & \multicolumn{2}{c}{24.75} & \multicolumn{2}{c|}{0.9271} & \multicolumn{2}{c}{17.57} & \multicolumn{2}{c|}{0.77} & \multicolumn{2}{c}{16.62} & \multicolumn{2}{c|}{0.52} & \multicolumn{2}{c}{13.84} & \multicolumn{2}{c}{0.43}  \\
\multicolumn{8}{c|}{(ICCV'17)AOD-Net\cite{aodnet}}                 & \multicolumn{2}{c}{20.51} & \multicolumn{2}{c|}{0.8164} & \multicolumn{2}{c}{24.14} & \multicolumn{2}{c|}{0.9203} & \multicolumn{2}{c}{15.03} & \multicolumn{2}{c|}{0.54} & \multicolumn{2}{c}{15.40} & \multicolumn{2}{c|}{0.57} & \multicolumn{2}{c}{13.14} & \multicolumn{2}{c}{0.41}  \\
\multicolumn{8}{c|}{(ICCV'19)GridDehazeNet\cite{griddehazenet}}           & \multicolumn{2}{c}{32.16} & \multicolumn{2}{c|}{0.9845} & \multicolumn{2}{c}{30.86} & \multicolumn{2}{c|}{0.9827} & \multicolumn{2}{c}{22.11} & \multicolumn{2}{c|}{0.71} & \multicolumn{2}{c}{13.80} & \multicolumn{2}{c|}{0.54} & \multicolumn{2}{c}{-}     & \multicolumn{2}{c}{-}    \\
\multicolumn{8}{c|}{(CVPR'20)MSBDN\cite{MSBDN}}                   & \multicolumn{2}{c}{33.67} & \multicolumn{2}{c|}{0.9856} & \multicolumn{2}{c}{33.48} & \multicolumn{2}{c|}{0.9824} & \multicolumn{2}{c}{24.36} & \multicolumn{2}{c|}{0.75} & \multicolumn{2}{c}{19.23} & \multicolumn{2}{c|}{0.71} & \multicolumn{2}{c}{15.37} & \multicolumn{2}{c}{0.49} \\
\multicolumn{8}{c|}{(AAAI'20)FFA-Net\cite{ffaNET}}                 & \multicolumn{2}{c}{36.39} & \multicolumn{2}{c|}{0.9894} & \multicolumn{2}{c}{33.57} & \multicolumn{2}{c|}{0.9842} & \multicolumn{2}{c}{22.12} & \multicolumn{2}{c|}{0.77} & \multicolumn{2}{c}{19.87} & \multicolumn{2}{c|}{0.69} & \multicolumn{2}{c}{14.39} & \multicolumn{2}{c}{0.45} \\
\multicolumn{8}{c|}{(CVPR'22)DeHamer\cite{dehamer}}                 & \multicolumn{2}{c}{36.63} & \multicolumn{2}{c|}{0.9881} & \multicolumn{2}{c}{35.18} & \multicolumn{2}{c|}{0.9860} & \multicolumn{2}{c}{24.64} & \multicolumn{2}{c|}{0.77} & \multicolumn{2}{c}{20.66} & \multicolumn{2}{c|}{0.68} & \multicolumn{2}{c}{16.62} & \multicolumn{2}{c}{0.56} \\
\multicolumn{8}{c|}{(ECCV'22)PMNet\cite{PMnet}}                   & \multicolumn{2}{c}{38.41} & \multicolumn{2}{c|}{0.9900} & \multicolumn{2}{c}{34.74} & \multicolumn{2}{c|}{0.9850} & \multicolumn{2}{c}{24.64} & \multicolumn{2}{c|}{0.83} & \multicolumn{2}{c}{20.42} & \multicolumn{2}{c|}{0.73} & \multicolumn{2}{c}{16.79} & \multicolumn{2}{c}{0.51} \\
\multicolumn{8}{c|}{(TIP'23)Dehazeformer\cite{dehazeformer}}          & \multicolumn{2}{c}{38.46} & \multicolumn{2}{c|}{0.9940} & \multicolumn{2}{c}{34.29} & \multicolumn{2}{c|}{0.9830} & \multicolumn{2}{c}{25.13} & \multicolumn{2}{c|}{0.77} & \multicolumn{2}{c}{19.11} & \multicolumn{2}{c|}{0.66} & \multicolumn{2}{c}{-}     & \multicolumn{2}{c}{-}    \\
\multicolumn{8}{c|}{(ICCV'23)FocalNet\cite{focalnet}}                & \multicolumn{2}{c}{40.82} & \multicolumn{2}{c|}{0.9960} & \multicolumn{2}{c}{37.71} & \multicolumn{2}{c|}{\underline{0.9950}} & \multicolumn{2}{c}{25.50} & \multicolumn{2}{c|}{\underline{0.94}} & \multicolumn{2}{c}{20.43} & \multicolumn{2}{c|}{0.79} & \multicolumn{2}{c}{17.07} & \multicolumn{2}{c}{0.63}\\
\multicolumn{8}{c|}{(CVPR'23)$C^{2}$PNet\cite{C2PNet}}                   & \multicolumn{2}{c}{42.56} & \multicolumn{2}{c|}{0.9954} & \multicolumn{2}{c}{36.68} & \multicolumn{2}{c|}{0.9900} & \multicolumn{2}{c}{-}     & \multicolumn{2}{c|}{-}    & \multicolumn{2}{c}{-}     & \multicolumn{2}{c|}{-}    & \multicolumn{2}{c}{16.88} & \multicolumn{2}{c}{0.57}    \\
\multicolumn{8}{c|}{(ICCV'23)MB-TaylorFormer\cite{MB-TaylorFormer}}                & \multicolumn{2}{c}{42.64} & \multicolumn{2}{c|}{0.9940} & \multicolumn{2}{c}{38.09} & \multicolumn{2}{c|}{0.9910} & \multicolumn{2}{c}{25.31} & \multicolumn{2}{c|}{0.78} & \multicolumn{2}{c}{-} & \multicolumn{2}{c|}{-} & \multicolumn{2}{c}{16.44} & \multicolumn{2}{c}{0.57} \\
\multicolumn{8}{c|}{(ECCV'24)SFHformer\cite{sfhformer}}                        & \multicolumn{2}{c}{\underline{43.03}}      & \multicolumn{2}{c|}{\underline{0.9966}}       & \multicolumn{2}{c}{\underline{38.83}}      & \multicolumn{2}{c|}{\textbf{0.9951}}       & \multicolumn{2}{c}{\underline{25.81}}      & \multicolumn{2}{c|}{\underline{0.94}}     & \multicolumn{2}{c}{\underline{20.73}}      & \multicolumn{2}{c|}{\underline{0.80}}     & \multicolumn{2}{c}{17.84}      & \multicolumn{2}{c}{\textbf{0.68}}     \\ \hline
\multicolumn{8}{c|}{(Ours)SWFormer-s}                        & \multicolumn{2}{c}{41.86}      & \multicolumn{2}{c|}{0.9962}       & \multicolumn{2}{c}{37.62}      & \multicolumn{2}{c|}{0.9945}       & \multicolumn{2}{c}{22.90}      & \multicolumn{2}{c|}{0.76}     & \multicolumn{2}{c}{19.17}      & \multicolumn{2}{c|}{0.70}     & \multicolumn{2}{c}{17.57}      & \multicolumn{2}{c}{\underline{0.65}}     \\
\multicolumn{8}{c|}{(Ours)SWFormer-m}                        & \multicolumn{2}{c}{42.54}      & \multicolumn{2}{c|}{0.9965}       & \multicolumn{2}{c}{38.57}      & \multicolumn{2}{c|}{\underline{0.9950}}       & \multicolumn{2}{c}{23.91}      & \multicolumn{2}{c|}{0.84}     & \multicolumn{2}{c}{18.44}      & \multicolumn{2}{c|}{0.72}     & \multicolumn{2}{c}{\underline{18.04}}      & \multicolumn{2}{c}{\textbf{0.68}}     \\
\multicolumn{8}{c|}{(Ours)SWFormer-l}                        & \multicolumn{2}{c}{\textbf{43.21}}      & \multicolumn{2}{c|}{\textbf{0.9968}}       & \multicolumn{2}{c}{\textbf{39.04}}      & \multicolumn{2}{c|}{\textbf{0.9951}}       & \multicolumn{2}{c}{\textbf{25.94}}      & \multicolumn{2}{c|}{\textbf{0.95}}     & \multicolumn{2}{c}{\textbf{20.86}}      & \multicolumn{2}{c|}{\textbf{0.82}}     & \multicolumn{2}{c}{\textbf{18.38}}      & \multicolumn{2}{c}{\textbf{0.68}}     \\ \hline
\end{tabular}}}
\vspace{1mm}
\end{minipage}

\begin{minipage}[c]{\textwidth}
\includegraphics[width=\linewidth]{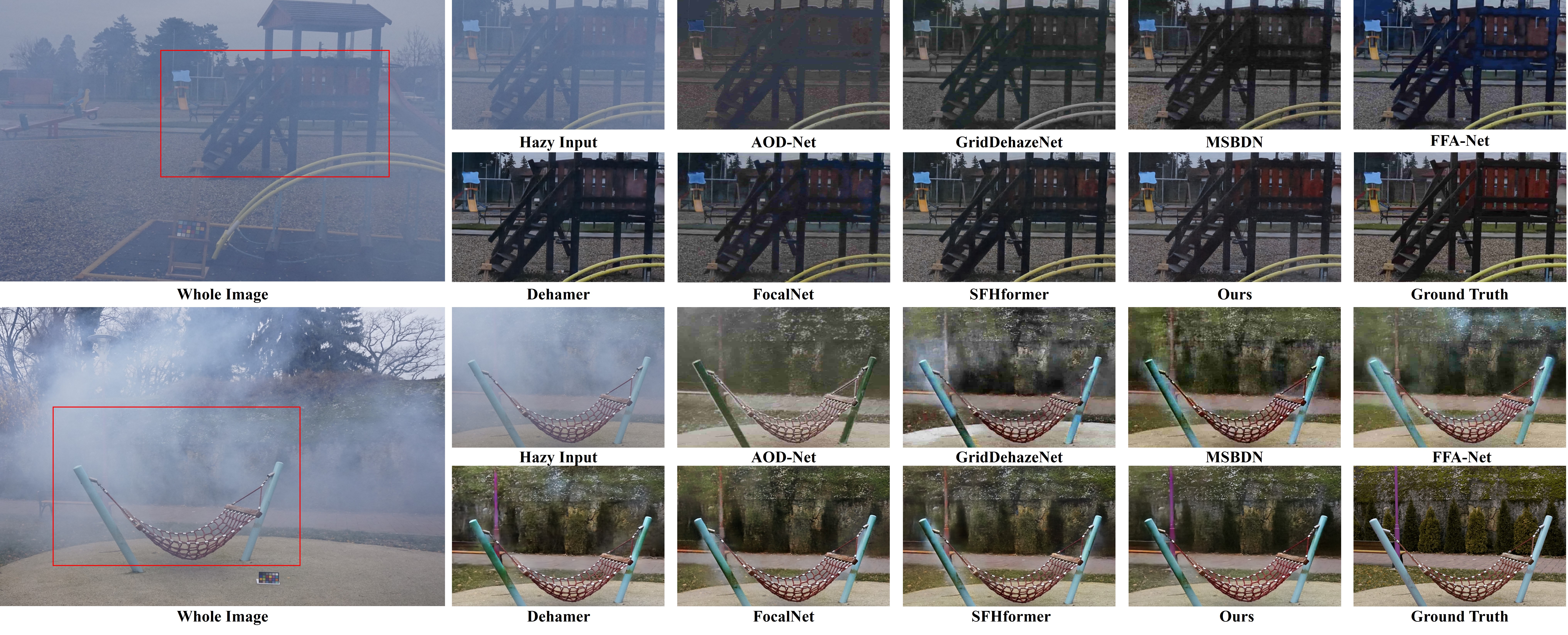}
\captionof{figure}{The quantitative evaluation results on synthetic and real-world image dehazing.} \label{fig:haze}
\end{minipage}
\end{table*}

We conduct dehazing experiments on five widely established public datasets, including two large-scale synthetic datasets, ITS \cite{reside} and OTS \cite{reside}, as well as three high-resolution real-world outdoor dehazing datasets: O-HAZE \cite{Ohaze}, NH-HAZE \cite{nhHAZE} and DENSE-HAZE \cite{denseHAZE}, which cover a range of dehazing scenarios such as outdoor, uneven and heavy haze removal. Tab.\ref{tab:haze} presents the quantitative comparison results for the dehazing task. Compared to the latest advanced methods, our proposed SWFormer consistently achieves the best restoration performance in terms of both PSNR and SSIM across all datasets. Notably, on the DENSE-HAZE dataset, our model outperforms SFHFormer \cite{sfhformer} by a 0.54 dB PSNR improvement, demonstrating its capability in handling extreme weather conditions, particularly heavy haze. Fig.\ref{fig:haze} illustrates the qualitative comparison results. Our method successes in delivering superior dehazing outcomes, whether for real-world outdoor dehazing or challenging uneven dehazing scenarios. A key strength of SWFormer lies in its remarkable performance in color correction. For instance, it effectively restores the natural color tones of red wooden planks and purple streetlights, ensuring more accurate and visually appealing results.

\subsection{Results on Motion Deblurring Task}

We perform motion blur removal experiments on two prominent large-scale benchmarks: GoPro\cite{GOPRO} and HIDE\cite{HIDE}. To assess the generalization capabilities of our approach, our model is trained exclusively on the GoPro dataset and then applied directly to the HIDE dataset. Tab.\ref{tab:motion} presents the quantitative comparison results for motion blur removal, underscoring the effectiveness of our method. In comparison to the latest state-of-the-art methods, our proposed SWFormer consistently outperforms existing models in restoration performance. Notably, on the HIDE dataset, our model achieves a substantial 0.41 dB PSNR improvement over SFHFormer\cite{sfhformer}, demonstrating its enhanced generalization ability when applied to previously unseen data. Fig.\ref{fig:motion-blur} illustrates the qualitative comparison results for motion blur removal. Our method stands out in its ability to recover fine high-frequency details, particularly in heavily degraded areas such as license plates and text, where precise clarity is paramount. In contrast, other methods introduce noticeable artifacts into their restored images, and in some instances, fail to adequately remove the blur.

\begin{table}[h]
\setlength{\abovecaptionskip}{0cm}
\setlength{\belowcaptionskip}{0cm}
\renewcommand{\arraystretch}{1}
\centering
\caption{Quantitative evaluations on motion deblurring.\label{tab:motion}}
\resizebox{0.48\textwidth}{!}{
\setlength{\tabcolsep}{0.85mm}{
\begin{tabular}{clllllll|clcl|clcl}
\hline
\multicolumn{8}{c|}{\multirow{2}{*}{Method}} & \multicolumn{4}{c|}{GoPro\cite{GOPRO}}                             & \multicolumn{4}{c}{HIDE\cite{HIDE}}                                                  \\ \cline{9-16} 
\multicolumn{8}{c|}{}  & \multicolumn{2}{c}{PSNR$\uparrow$}  & \multicolumn{2}{c|}{SSIM$\uparrow$}  & \multicolumn{2}{c}{PSNR$\uparrow$}  & \multicolumn{2}{c}{SSIM$\uparrow$}    \\ \hline
% \multicolumn{8}{c|}{(CVPR'20)DBGAN\cite{DBGAN}}                   & \multicolumn{2}{c}{31.10} & \multicolumn{2}{c|}{0.942} & \multicolumn{2}{c}{28.94} & \multicolumn{2}{c}{0.915}        \\
% \multicolumn{8}{c|}{(ECCV'20)MT-RNN\cite{MT-RNN}}                  & \multicolumn{2}{c}{31.15} & \multicolumn{2}{c|}{0.945} & \multicolumn{2}{c}{29.15} & \multicolumn{2}{c}{0.918}        \\
% \multicolumn{8}{c|}{(CVPR'19)DMPHN\cite{DMPHN}}                   & \multicolumn{2}{c}{31.20} & \multicolumn{2}{c|}{0.940} & \multicolumn{2}{c}{29.09} & \multicolumn{2}{c}{0.924}      \\
% \multicolumn{8}{c|}{(ICCV'21)SPAIR\cite{SPAIR}}                   & \multicolumn{2}{c}{32.06} & \multicolumn{2}{c|}{0.953} & \multicolumn{2}{c}{30.29} & \multicolumn{2}{c}{0.931}       \\
\multicolumn{8}{c|}{(ICCV'21)MIMO-UNet+\cite{MIMO-UNet}}              & \multicolumn{2}{c}{32.45} & \multicolumn{2}{c|}{0.957} & \multicolumn{2}{c}{29.99} & \multicolumn{2}{c}{0.930}       \\
\multicolumn{8}{c|}{(CVPR'21)MPRNet\cite{MPRNet}}                  & \multicolumn{2}{c}{32.66} & \multicolumn{2}{c|}{0.959} & \multicolumn{2}{c}{30.96} & \multicolumn{2}{c}{0.939}       \\
\multicolumn{8}{c|}{(CVPR'22)Restormer\cite{Restormer}}               & \multicolumn{2}{c}{32.92} & \multicolumn{2}{c|}{0.961} & \multicolumn{2}{c}{31.22} & \multicolumn{2}{c}{0.942}        \\
\multicolumn{8}{c|}{(ECCV'22)Stripformer\cite{stripformer}}             & \multicolumn{2}{c}{33.08} & \multicolumn{2}{c|}{0.962} & \multicolumn{2}{c}{31.03} & \multicolumn{2}{c}{0.940}         \\
\multicolumn{8}{c|}{(ECCV'22)MPRNet-local\cite{MPRNet-local}}            & \multicolumn{2}{c}{33.31} & \multicolumn{2}{c|}{0.964} & \multicolumn{2}{c}{31.19} & \multicolumn{2}{c}{0.942}       \\
\multicolumn{8}{c|}{(ECCV'22)Restormer-local\cite{MPRNet-local}}         & \multicolumn{2}{c}{33.57} & \multicolumn{2}{c|}{0.966} & \multicolumn{2}{c}{31.49} & \multicolumn{2}{c}{0.945}        \\
\multicolumn{8}{c|}{(ECCV'22)NAFNet\cite{NAFNet}}                  & \multicolumn{2}{c}{33.71} & \multicolumn{2}{c|}{0.967} & \multicolumn{2}{c}{31.31} & \multicolumn{2}{c}{0.943}    \\
\multicolumn{8}{c|}{(ICLR'23)SFNet\cite{sfnet}}                   & \multicolumn{2}{c}{33.27} & \multicolumn{2}{c|}{0.963} & \multicolumn{2}{c}{31.10} & \multicolumn{2}{c}{0.941} \\
% \multicolumn{8}{c|}{(ICML'23)IRNeXt\cite{IRNeXt}}                  & \multicolumn{2}{c}{33.16} & \multicolumn{2}{c|}{0.962} & \multicolumn{2}{c}{-}      & \multicolumn{2}{c}{-}     \\
\multicolumn{8}{c|}{(ICCV'23)icDPMs-SA\cite{icDPMs-SA}}               & \multicolumn{2}{c}{33.20} & \multicolumn{2}{c|}{0.963} & \multicolumn{2}{c}{30.96} & \multicolumn{2}{c}{0.938}      \\
\multicolumn{8}{c|}{(CVPR'23)GRL\cite{GRL-B}}                   & \multicolumn{2}{c}{33.93} & \multicolumn{2}{c|}{\underline{0.968}} & \multicolumn{2}{c}{31.62} & \multicolumn{2}{c}{0.947}       \\ 
\multicolumn{8}{c|}{(ECCV'24)X-Restormer\cite{x-restormer}}         & \multicolumn{2}{c}{33.44} & \multicolumn{2}{c|}{0.946}      & \multicolumn{2}{c}{\underline{31.76}} & \multicolumn{2}{c}{0.930}            \\
\multicolumn{8}{c|}{(CVPR'24)MISC Filter\cite{miscfilter}}         & \multicolumn{2}{c}{\underline{34.10}} & \multicolumn{2}{c|}{\textbf{0.969}}      & \multicolumn{2}{c}{31.66} & \multicolumn{2}{c}{0.946}            \\
\multicolumn{8}{c|}{(ECCV'24)SFHformer\cite{sfhformer}}         & \multicolumn{2}{c}{34.01} & \multicolumn{2}{c|}{\textbf{0.969}}      & \multicolumn{2}{c}{31.66} & \multicolumn{2}{c}{\underline{0.948}}            \\ \hline
\multicolumn{8}{c|}{(Ours)SWFormer-s}         & \multicolumn{2}{c}{33.55} & \multicolumn{2}{c|}{0.963}      & \multicolumn{2}{c}{31.27} & \multicolumn{2}{c}{0.943}            \\
\multicolumn{8}{c|}{(Ours)SWFormer-m}         & \multicolumn{2}{c}{33.94} & \multicolumn{2}{c|}{0.967}      & \multicolumn{2}{c}{31.72} & \multicolumn{2}{c}{\underline{0.948}}            \\
\multicolumn{8}{c|}{(Ours)SWFormer-l}         & \multicolumn{2}{c}{\textbf{34.21}} & \multicolumn{2}{c|}{\textbf{0.969}}      & \multicolumn{2}{c}{\textbf{32.07}} & \multicolumn{2}{c}{\textbf{0.950}}            \\ \hline
\end{tabular}}}
\vspace{-3mm}
\end{table}

\begin{table*}[!t]
\setlength{\abovecaptionskip}{0cm}
\setlength{\belowcaptionskip}{0cm}
\centering
\begin{minipage}[c]{\textwidth}
\includegraphics[width=\linewidth]{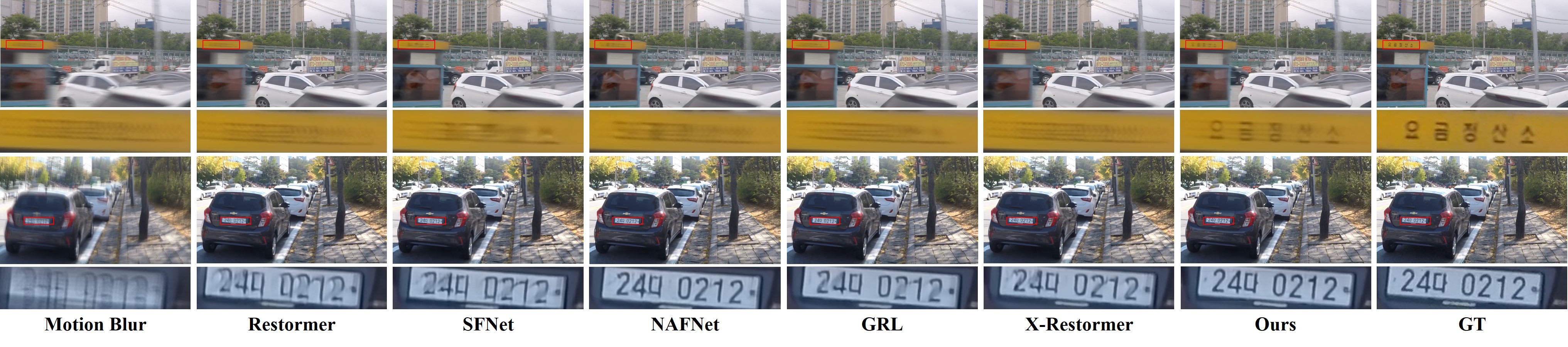}
\captionof{figure}{The quantitative evaluation results on motion deblurring.\label{fig:motion-blur}}
\end{minipage}
\begin{minipage}[c]{\textwidth}
\includegraphics[width=\linewidth]{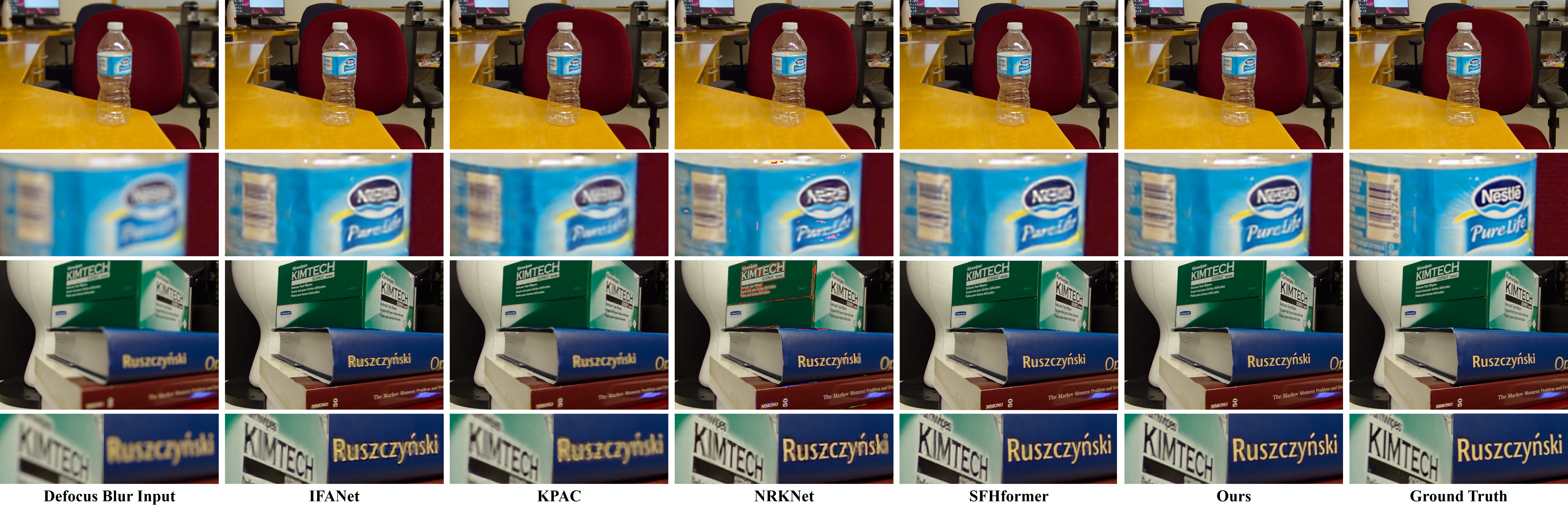}
\captionof{figure}{The quantitative evaluation results on defocus deblurring.} \label{fig:defocus blur}
\end{minipage}
\begin{minipage}[c]{\textwidth}
\includegraphics[width=\linewidth]{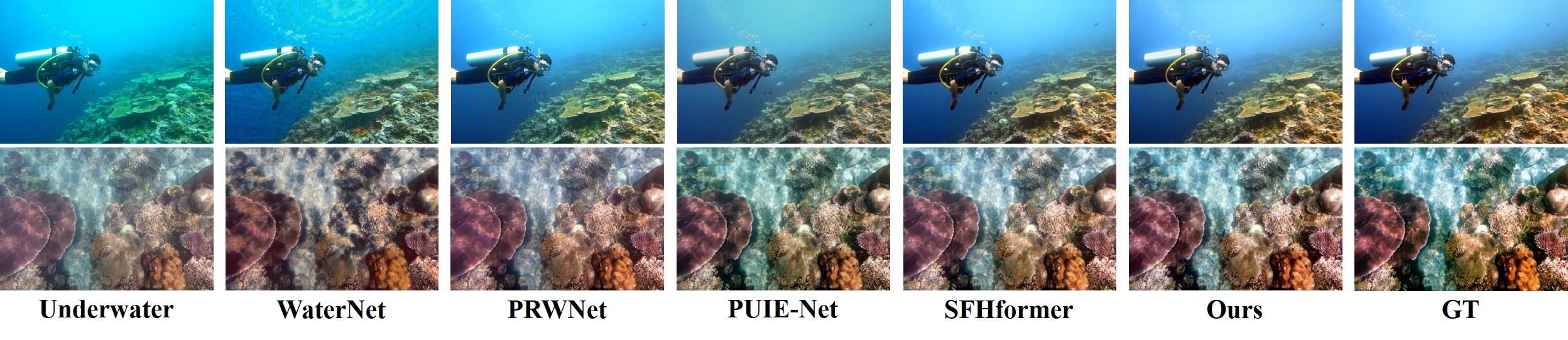}
\captionof{figure}{The quantitative evaluation results on underwater enhancement.} \label{fig:underwater}
\end{minipage}
\vspace{-4mm}
\end{table*}

\subsection{Results on Defocus Deblurring Task}
\begin{table}[h]
\setlength{\abovecaptionskip}{0cm}
\setlength{\belowcaptionskip}{0cm}
\renewcommand{\arraystretch}{1}
\centering
\caption{Quantitative evaluations on defocus blur.\label{tab:defocus blur}}
\resizebox{0.48\textwidth}{!}{
\setlength{\tabcolsep}{1.1mm}{
\begin{tabular}{cl|ccc|ccc}
\hline
\multicolumn{2}{c|}{\multirow{2}{*}{Method}} & \multicolumn{3}{c|}{DPDD\cite{DPDD}} & \multicolumn{3}{c}{LFDOF\cite{aifnet}} \\ \cline{3-8} 
\multicolumn{2}{c|}{}                        & PSNR$\uparrow$    & SSIM$\uparrow$   & LPIPS$\downarrow$  & PSNR$\uparrow$    & SSIM$\uparrow$   & LPIPS$\downarrow$  \\ \hline
\multicolumn{2}{c|}{(TCI'21)AIFNet\cite{aifnet}}                  & 24.213  & 0.742  & 0.309  & 29.677  & 0.884  & 0.202  \\
\multicolumn{2}{c|}{(WACV'22)MDP\cite{mdp}}                     & 25.347  & 0.763  & 0.268  & 28.069  & 0.834  & 0.185  \\
\multicolumn{2}{c|}{(CVPR'21)KPAC\cite{KPAC}}                    & 25.221  & 0.774  & 0.226  & 28.942  & 0.857  & 0.174  \\
\multicolumn{2}{c|}{(CVPR'21)IFANet\cite{ifanet}}                  & 25.366  & 0.789  & 0.217  & 29.787  & 0.872  & 0.154  \\
\multicolumn{2}{c|}{(NIPS'21)GKMNet\cite{gkmnet}}                  & 25.468  & 0.789  & 0.219  & 29.081  & 0.867  & 0.171  \\
\multicolumn{2}{c|}{(CVPR'22)DRBNet\cite{drbnet}}                 & 25.485  & 0.792  & 0.254  & 30.253  & 0.883  & 0.147  \\
\multicolumn{2}{c|}{(CVPR'23)NRKNet\cite{NRKNet}}                  & 26.109  & 0.810  & \textbf{0.210}  & 30.481  & 0.884  & 0.147  \\
\multicolumn{2}{c|}{(ECCV'24)SFHformer\cite{sfhformer}}               & 26.118  & 0.807  & 0.222  & 30.623       & 0.886      & 0.146      \\
\multicolumn{2}{c|}{(TPAMI'24)GGKMNet\cite{ggkmnet}}                 & \textbf{26.272}  & 0.810  & \underline{0.215}  & 30.552  & 0.886  & 0.154  \\ \hline
\multicolumn{2}{c|}{(Ours)SWFormer-s}              & 26.126  & 0.810  & 0.220  & 30.563  & 0.884  & 0.142  \\
\multicolumn{2}{c|}{(Ours)SWFormer-m}              & 26.184  & \underline{0.817}  & 0.219  & \underline{30.954}  & \underline{0.892}  & \underline{0.141}  \\
\multicolumn{2}{c|}{(Ours)SWFormer-l}              & \underline{26.220}  & \textbf{0.819}  & 0.221  & \textbf{31.270}   & \textbf{0.898}  & \textbf{0.133}  \\ \hline
\end{tabular}}}
\vspace{-5mm}
\end{table}

Building on the conference version \cite{sfhformer}, we perform defocus blur removal experiments on two widely used real-world public datasets: DPDD\cite{DPDD} and the newly introduced large-scale LFDOF\cite{aifnet} dataset. Tab.\ref{tab:defocus blur} presents the quantitative comparison results for defocus blur removal. Our proposed SWFormer demonstrates exceptional performance across all datasets. Notably, on the LFDOF dataset, our model achieves the best performance in terms of PSNR, SSIM, and LPIPS. Fig.\ref{fig:defocus blur} presents the qualitative comparison results. Visually, our method recovers sharper high-frequency details, crucial for accurately restoring fine structures such as text and edges. In contrast, NRKNet \cite{NRKNet} suffers from numerical overflow issues, resulting in degraded performance and less precise restoration.

\begin{table}[h]
\setlength{\abovecaptionskip}{0cm}
\setlength{\belowcaptionskip}{0cm}
\renewcommand{\arraystretch}{1}
\centering
\caption{Evaluations on underwater enhancement.\label{tab:water}}
\resizebox{0.48\textwidth}{!}{
\setlength{\tabcolsep}{0.8mm}{
\begin{tabular}{clllllll|clclcl|clclcl}
\hline
\multicolumn{8}{c|}{\multirow{2}{*}{Method}} & \multicolumn{6}{c|}{L-400\cite{U-shapeTranslsui}}                                                           & \multicolumn{6}{c}{U-90\cite{waternetuieb}}                                                                                   \\ \cline{9-20} 
\multicolumn{8}{c|}{}                        & \multicolumn{2}{c}{PSNR$\uparrow$}  & \multicolumn{2}{c}{SSIM$\uparrow$}   & \multicolumn{2}{c|}{UCIQE$\uparrow$}  & \multicolumn{2}{c}{PSNR$\uparrow$}  & \multicolumn{2}{c}{SSIM$\uparrow$}   & \multicolumn{2}{c}{UCIQE$\uparrow$}   \\ \hline
\multicolumn{8}{c|}{(TIP'20)WaterNet\cite{waternetuieb}}        & \multicolumn{2}{c}{23.38} & \multicolumn{2}{c}{0.9152} & \multicolumn{2}{c|}{0.5729} & \multicolumn{2}{c}{16.31} & \multicolumn{2}{c}{0.7970} & \multicolumn{2}{c}{0.5777}  \\
\multicolumn{8}{c|}{(ICCV'21)PRWNet\cite{PRWNet}}        & \multicolumn{2}{c}{27.83} & \multicolumn{2}{c}{0.9268} & \multicolumn{2}{c|}{0.5801} & \multicolumn{2}{c}{20.79} & \multicolumn{2}{c}{0.8231} & \multicolumn{2}{c}{0.5830} \\
\multicolumn{8}{c|}{(AAAI'21)Shallow-UWNet\cite{Shallow-UWNet}}  & \multicolumn{2}{c}{20.56} & \multicolumn{2}{c}{0.7675} & \multicolumn{2}{c|}{0.5517} & \multicolumn{2}{c}{18.28} & \multicolumn{2}{c}{0.8553} & \multicolumn{2}{c}{0.5517}  \\
\multicolumn{8}{c|}{(TIP'23)U-shape Trans\cite{U-shapeTranslsui}}  & \multicolumn{2}{c}{24.16} & \multicolumn{2}{c}{0.9184} & \multicolumn{2}{c|}{0.5871} & \multicolumn{2}{c}{21.25} & \multicolumn{2}{c}{0.8432} & \multicolumn{2}{c}{0.5882}  \\
\multicolumn{8}{c|}{(ECCV'22)PUIE-Net\cite{PUIE-Net}}       & \multicolumn{2}{c}{21.28} & \multicolumn{2}{c}{0.8615} & \multicolumn{2}{c|}{0.5887} & \multicolumn{2}{c}{21.38} & \multicolumn{2}{c}{0.8821} & \multicolumn{2}{c}{0.5887} \\ 
\multicolumn{8}{c|}{(TGRS'22)URSCT-SESR\cite{URSCT-SESR}}     & \multicolumn{2}{c}{29.31} & \multicolumn{2}{c}{0.9318} & \multicolumn{2}{c|}{0.5890} & \multicolumn{2}{c}{22.72} & \multicolumn{2}{c}{0.9108} & \multicolumn{2}{c}{\textbf{0.6140}} \\
\multicolumn{8}{c|}{(ECCV'24)SFHformer\cite{sfhformer}}                        & \multicolumn{2}{c}{30.18}      & \multicolumn{2}{c}{\underline{0.9449}}       & \multicolumn{2}{c|}{\textbf{0.5943}}       & \multicolumn{2}{c}{\underline{23.54}}      & \multicolumn{2}{c}{\underline{0.9177}}       & \multicolumn{2}{c}{\underline{0.6020}}      \\ 
\hline
\multicolumn{8}{c|}{(Ours)SWFormer-s}                        & \multicolumn{2}{c}{30.16}      & \multicolumn{2}{c}{0.9422}       & \multicolumn{2}{c|}{0.5907}       & \multicolumn{2}{c}{23.10}      & \multicolumn{2}{c}{0.9083}       & \multicolumn{2}{c}{0.5990}      \\
\multicolumn{8}{c|}{(Ours)SWFormer-m}                        & \multicolumn{2}{c}{\underline{30.49}}      & \multicolumn{2}{c}{0.9447}       & \multicolumn{2}{c|}{0.5921}       & \multicolumn{2}{c}{23.46}      & \multicolumn{2}{c}{0.9138}       & \multicolumn{2}{c}{0.5975}      \\
\multicolumn{8}{c|}{(Ours)SWFormer-l}                        & \multicolumn{2}{c}{\textbf{30.56}}      & \multicolumn{2}{c}{\textbf{0.9467}}       & \multicolumn{2}{c|}{\underline{0.5932}}       & \multicolumn{2}{c}{\textbf{23.69}}      & \multicolumn{2}{c}{\textbf{0.9187}}       & \multicolumn{2}{c}{0.6019}      \\\hline
\end{tabular}}}
\vspace{-5mm}
\end{table}
\subsection{Results on Underwater Image Enhancement Task}

We conduct underwater enhancement experiments on two real-world benchmarks: UIEB \cite{waternetuieb} and LSUI \cite{U-shapeTranslsui}. Tab.\ref{tab:water} presents the quantitative comparison results, using PSNR and SSIM as reference-based metrics, and UCIQE as a no-reference metric. Compared to the latest methods, SWFormer demonstrates competitive performance, achieving a 0.38 dB PSNR improvement over SFHFormer\cite{sfhformer} on the large-scale LSUI dataset. Fig.\ref{fig:underwater} presents the qualitative comparison results. The images restored by our model are closer to the ground truth, excelling in color correction and preserving natural color tones better than other methods in underwater.

\begin{table*}[!t]
\setlength{\abovecaptionskip}{0cm}
\setlength{\belowcaptionskip}{0cm}
\centering
\begin{minipage}[c]{\textwidth}
\includegraphics[width=\linewidth]{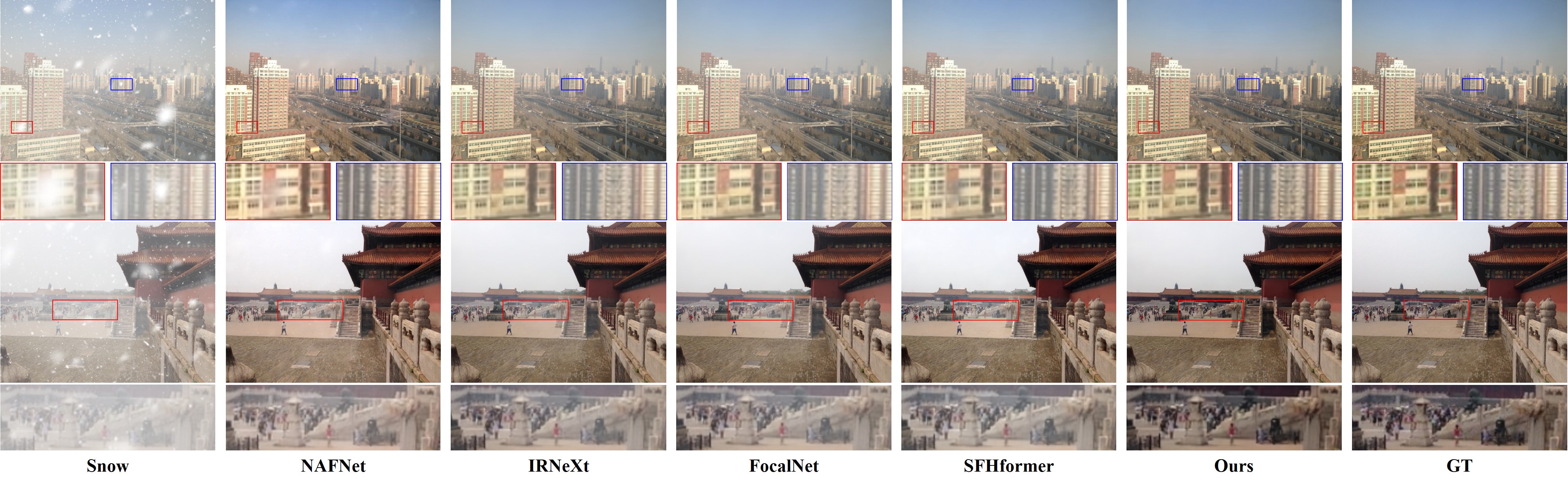}
\captionof{figure}{The quantitative evaluation results on image desnowing.\label{fig:snow}}
\end{minipage}
\begin{minipage}[c]{\textwidth}
\includegraphics[width=\linewidth]{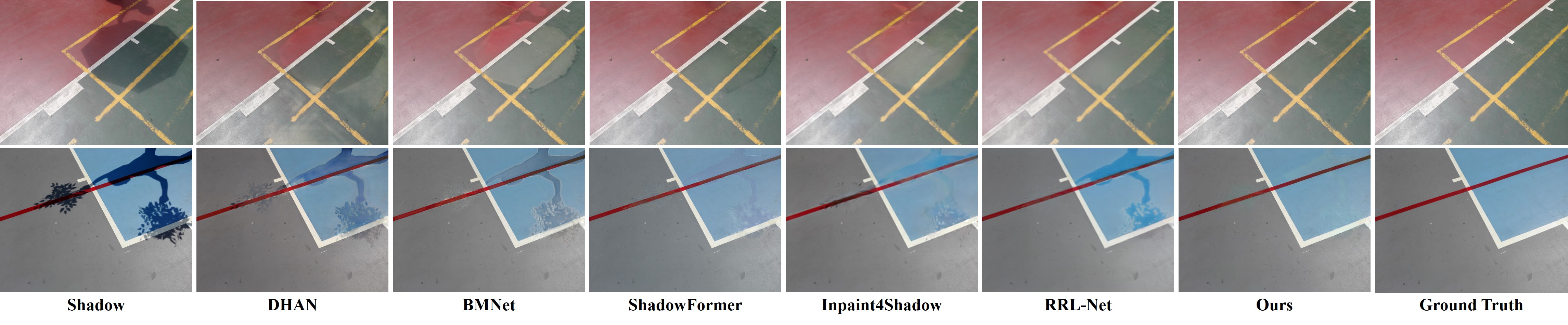}
\captionof{figure}{The quantitative evaluation results on shadow removal.} \label{fig:shadow}
\end{minipage}
\end{table*}

\begin{table}[h]
\setlength{\abovecaptionskip}{0cm}
\setlength{\belowcaptionskip}{0cm}
\centering
\renewcommand{\arraystretch}{1}
\caption{Quantitative evaluations on image desnowing.}\label{tab:snow}
\resizebox{0.45\textwidth}{!}{
\setlength{\tabcolsep}{0.8mm}{
\begin{tabular}{clllllll|clcl|clcl|clcl}
\hline
\multicolumn{8}{c|}{\multirow{2}{*}{Method}} & \multicolumn{4}{c|}{\makecell{CSD\cite{HDCW-NetCSD}}}            & \multicolumn{4}{c|}{\makecell{SRRS\cite{JSTASRSRRS}}}           & \multicolumn{4}{c}{\makecell{Snow100K\cite{desnownesnow100k}}}                 \\ \cline{9-20} 
\multicolumn{8}{c|}{}                        & \multicolumn{2}{c}{PSNR$\uparrow$}  & \multicolumn{2}{c|}{SSIM$\uparrow$} & \multicolumn{2}{c}{PSNR$\uparrow$}  & \multicolumn{2}{c|}{SSIM$\uparrow$} & \multicolumn{2}{c}{PSNR$\uparrow$}  & \multicolumn{2}{c}{SSIM$\uparrow$} \\ \hline
% \multicolumn{8}{c|}{(TIP'18)DesnowNet\cite{desnownesnow100k}}       & \multicolumn{2}{c}{20.13} & \multicolumn{2}{c|}{0.81} & \multicolumn{2}{c}{20.38} & \multicolumn{2}{c|}{0.84} & \multicolumn{2}{c}{30.50}   & \multicolumn{2}{c}{0.94}      \\
% \multicolumn{8}{c|}{(CVPR'18)CycleGAN\cite{CycleGAN}}       & \multicolumn{2}{c}{20.98} & \multicolumn{2}{c|}{0.80} & \multicolumn{2}{c}{20.21} & \multicolumn{2}{c|}{0.74} & \multicolumn{2}{c}{26.81} & \multicolumn{2}{c}{0.89}  \\
\multicolumn{8}{c|}{(CVPR'22)ALL in One\cite{allinone}}     & \multicolumn{2}{c}{26.31} & \multicolumn{2}{c|}{0.87} & \multicolumn{2}{c}{24.98} & \multicolumn{2}{c|}{0.88} & \multicolumn{2}{c}{26.07} & \multicolumn{2}{c}{0.88} \\
\multicolumn{8}{c|}{(ECCV'20)JSTASR\cite{JSTASRSRRS}}         & \multicolumn{2}{c}{27.96} & \multicolumn{2}{c|}{0.88} & \multicolumn{2}{c}{25.82} & \multicolumn{2}{c|}{0.89} & \multicolumn{2}{c}{23.12} & \multicolumn{2}{c}{0.86}    \\
\multicolumn{8}{c|}{(ICCV'21)HDCW-Net\cite{HDCW-NetCSD}}       & \multicolumn{2}{c}{29.06} & \multicolumn{2}{c|}{0.91} & \multicolumn{2}{c}{27.78} & \multicolumn{2}{c|}{0.92} & \multicolumn{2}{c}{31.54} & \multicolumn{2}{c}{0.95}  \\
\multicolumn{8}{c|}{(CVPR'22)TransWeather\cite{transweather}}   & \multicolumn{2}{c}{31.76} & \multicolumn{2}{c|}{0.93} & \multicolumn{2}{c}{28.29} & \multicolumn{2}{c|}{0.92} & \multicolumn{2}{c}{31.82} & \multicolumn{2}{c}{0.93}  \\
\multicolumn{8}{c|}{(ECCV'22)NAFNet\cite{NAFNet}}         & \multicolumn{2}{c}{33.13} & \multicolumn{2}{c|}{0.96} & \multicolumn{2}{c}{29.72} & \multicolumn{2}{c|}{\underline{0.94}} & \multicolumn{2}{c}{32.41} & \multicolumn{2}{c}{\underline{0.95}} \\
\multicolumn{8}{c|}{(ICCV'23)FocalNet\cite{focalnet}}       & \multicolumn{2}{c}{37.18} & \multicolumn{2}{c|}{\textbf{0.99}} & \multicolumn{2}{c}{31.34} & \multicolumn{2}{c|}{\textbf{0.98}} & \multicolumn{2}{c}{33.53} & \multicolumn{2}{c}{\underline{0.95}}  \\
\multicolumn{8}{c|}{(ICML'23)IRNeXt\cite{IRNeXt}}       & \multicolumn{2}{c}{37.29} & \multicolumn{2}{c|}{\textbf{0.99}} & \multicolumn{2}{c}{31.91} & \multicolumn{2}{c|}{\textbf{0.98}} & \multicolumn{2}{c}{33.61} & \multicolumn{2}{c}{\underline{0.95}} \\
\multicolumn{8}{c|}{(ECCV'24)SFHformer\cite{sfhformer}}                        & \multicolumn{2}{c}{\underline{37.45}}      & \multicolumn{2}{c|}{\textbf{0.99}}     & \multicolumn{2}{c}{32.39}      & \multicolumn{2}{c|}{\textbf{0.98}}     & \multicolumn{2}{c}{33.94}      & \multicolumn{2}{c}{\underline{0.95}}        \\ \hline
\multicolumn{8}{c|}{(Ours)SWFormer-s}                        & \multicolumn{2}{c}{36.85}      & \multicolumn{2}{c|}{\underline{0.98}}     & \multicolumn{2}{c}{31.83}      & \multicolumn{2}{c|}{\textbf{0.98}}     & \multicolumn{2}{c}{33.82}      & \multicolumn{2}{c}{\underline{0.95}}        \\
\multicolumn{8}{c|}{(Ours)SWFormer-m}                        & \multicolumn{2}{c}{37.43}      & \multicolumn{2}{c|}{\textbf{0.99}}     & \multicolumn{2}{c}{\underline{32.55}}      & \multicolumn{2}{c|}{\textbf{0.98}}     & \multicolumn{2}{c}{\underline{34.21}}      & \multicolumn{2}{c}{\textbf{0.96}}        \\
\multicolumn{8}{c|}{(Ours)SWFormer-l}                        & \multicolumn{2}{c}{\textbf{37.76}}      & \multicolumn{2}{c|}{\textbf{0.99}}     & \multicolumn{2}{c}{\textbf{32.81}}      & \multicolumn{2}{c|}{\textbf{0.98}}     & \multicolumn{2}{c}{\textbf{34.55}}      & \multicolumn{2}{c}{\textbf{0.96}}        \\
\hline
\end{tabular}}}
\end{table}

\subsection{Results on Image Desnowing Task}

We conduct snow removal experiments on three widely used large-scale public datasets: CSD\cite{HDCW-NetCSD}, SRRS\cite{JSTASRSRRS} and Snow100K\cite{desnownesnow100k}. Tab.\ref{tab:snow} presents the quantitative comparison results, where SWFormer consistently surpasses state-of-the-art methods, achieving the best performance in snow removal across all datasets. Notably, it shows an average PSNR improvement of 0.4 dB over SFHFormer\cite{sfhformer}. Fig.\ref{fig:snow} illustrates the qualitative results. Our model excels at recovering heavily snow-covered areas and effectively eliminating snow from distant views, delivering visually natural and superior outcomes. 

\subsection{Results on Shadow Removal Task}
\begin{table}[h]
\setlength{\abovecaptionskip}{0cm}
\setlength{\belowcaptionskip}{0cm}
\renewcommand{\arraystretch}{1}
\centering
\caption{Quantitative evaluations on shadow removal.\label{tab:shadow}}
\resizebox{0.48\textwidth}{!}{
\setlength{\tabcolsep}{1mm}{
\begin{tabular}{cl|ccc|ccc}
\hline
\multicolumn{2}{c|}{\multirow{2}{*}{Method}} & \multicolumn{3}{c|}{Shadow Region (S)} & \multicolumn{3}{c}{Whole Image (ALL)} \\ \cline{3-8} 
\multicolumn{2}{c|}{}                        & PSNR$\uparrow$        & SSIM$\uparrow$        & RMSE$\downarrow$       & PSNR$\uparrow$        & SSIM$\uparrow$        & RMSE$\downarrow$      \\ \hline
\multicolumn{2}{c|}{(AAAI'20)DHAN\cite{DHAN}}                    & 33.08       & 0.988       & 9.49       & 25.78       & 0.958       & 7.74      \\
% \multicolumn{2}{c|}{G2R}                     & 33.88       & 0.978       & 8.71       & 30.85       & 0.946       & 3.78      \\
\multicolumn{2}{c|}{(CVPR'21)DC-ShadowNet\cite{DC-ShadowNet}}            & 32.20       & 0.977       & 10.83      & 29.17       & 0.939       & 4.70      \\
\multicolumn{2}{c|}{(CVPR'21)AutoExposure\cite{AutoExposure}}            & 36.02       & 0.976       & 6.67       & 29.28       & 0.847       & 4.28      \\
\multicolumn{2}{c|}{(CVPR'22)BMNet\cite{BMNet}}                   & 38.17       & \underline{0.991}       & 5.72       & 34.34       & 0.974       & 2.93      \\
\multicolumn{2}{c|}{(AAAI'23)ShadowFormer\cite{ShadowFormer}}            & 39.67       & \textbf{0.992}       & 5.21       & 35.46       & 0.973       & 2.80      \\
\multicolumn{2}{c|}{(CVPR'23)ShadowDiffusion\cite{shadowdiffusion}}         & 39.82       & -           & 4.90       & 35.72       & -           & 2.70      \\
\multicolumn{2}{c|}{(ICCV'23)Inpaint4Shadow\cite{inpaint4shadow}}          & 38.46       & 0.989       & 5.93       & 34.14       & 0.960       & 3.39      \\
\multicolumn{2}{c|}{(AAAI'24)RRL-Net\cite{rrlnet}}                 & 38.04       & 0.990       & 5.69       & 34.96       & 0.968       & 2.87      \\ \hline
\multicolumn{2}{c|}{(Ours)SWFormer-s}              & 40.16       & \textbf{0.992}       & 5.04       & 35.68       & \underline{0.975}       & \underline{2.69}      \\
\multicolumn{2}{c|}{(Ours)SWFormer-m}              & \underline{40.37}       & \textbf{0.992}       & \underline{4.94}       & \textbf{35.85}       & \textbf{0.976}      & \textbf{2.66}      \\
\multicolumn{2}{c|}{(Ours)SWFormer-l}              & \textbf{40.45}       & \textbf{0.992}       & \textbf{4.86}       & \underline{35.79}       & \textbf{0.976}       & \textbf{2.66}      \\ \hline
\end{tabular}}}
\end{table}

For the shadow removal task, we conduct experiments using the AISTD\cite{aistd} real-world public dataset. Tab.\ref{tab:shadow} presents the quantitative comparison results, evaluating the restoration performance from two perspectives: the shadow region and the entire image. We use PSNR, SSIM and RMSE as evaluation metrics. Compared to the latest state-of-the-art methods, our proposed SWFormer achieves the highest performance. Notably, in the shadow region, our model surpasses the second-best diffusion-based method, ShadowDiffusion \cite{shadowdiffusion}, with a 0.63 dB PSNR improvement. Fig.\ref{fig:shadow} illustrates the qualitative comparison results. It is evident that our method provides the most comprehensive shadow removal, delivering the most natural and seamless visual results, without leaving any noticeable traces of shadow removal.

\subsection{Results on Remote Sensing Cloud Removal Task}
\begin{table}[h]
\setlength{\abovecaptionskip}{0cm}
\setlength{\belowcaptionskip}{0cm}
\renewcommand{\arraystretch}{1}
\centering
\begin{minipage}[c]{0.48\textwidth}
\captionof{table}{Quantitative evaluations on cloud removal.\label{tab:cloud}}
\resizebox{\textwidth}{!}{
\setlength{\tabcolsep}{1mm}{
\begin{tabular}{cl|ccc|ccc}
\hline
\multicolumn{2}{c|}{\multirow{2}{*}{Method}} & \multicolumn{3}{c|}{CUHK-CR1\cite{de-msda_memory}} & \multicolumn{3}{c}{CUHK-CR2\cite{de-msda_memory}} \\ \cline{3-8} 
\multicolumn{2}{c|}{}                        & PSNR$\uparrow$     & SSIM$\uparrow$     & LPIPS$\downarrow$   & PSNR$\uparrow$     & SSIM$\uparrow$    & LPIPS$\downarrow$   \\ \hline
\multicolumn{2}{c|}{(ACCV'22)CVAE\cite{CVAE}}                    & 24.252   & 0.7252   & 0.1075  & 22.631   & 0.6302  & 0.0489  \\
\multicolumn{2}{c|}{(ArXiV'23)MemoryNet\cite{memorynet}}               & 26.073   & 0.7741   & 0.0315  & 24.224   & 0.6838  & 0.0403  \\
\multicolumn{2}{c|}{(GRSL'22)MSDA-CR\cite{msda-cr}}                 & 25.435   & 0.7483   & 0.0374  & 23.755   & 0.6661  & 0.0433  \\
\multicolumn{2}{c|}{(TGRS'24)DE-MemoryNet\cite{de-msda_memory}}            & 26.183   & 0.7746   & 0.0290  & 24.348   & 0.6843  & 0.0369  \\
\multicolumn{2}{c|}{(TGRS'24)DE-MSDA\cite{de-msda_memory}}                 & 25.739   & 0.7592   & 0.0321  & 23.968   & 0.6737  & 0.0372  \\
\multicolumn{2}{c|}{(CVPR'25)EMRDM\cite{emrdm}}                   & 27.281   & 0.8007   & \textbf{0.0218}  & 24.594   & 0.6951  & \textbf{0.0301}  \\ \hline
\multicolumn{2}{c|}{(Ours)SWFormer-s}              & 27.213   & 0.8019   & 0.0267  & 25.520   & 0.7261  & 0.0351  \\
\multicolumn{2}{c|}{(Ours)SWFormer-m}              & \underline{27.380}   & \underline{0.8100}   & \underline{0.0241}  & \underline{25.706}   & \underline{0.7417}  & \underline{0.0326}  \\
\multicolumn{2}{c|}{(Ours)SWFormer-l}              & \textbf{27.425}   & \textbf{0.8126}   & 0.0242  & \textbf{25.737}   & \textbf{0.7443}  & 0.0329  \\ \hline
\end{tabular}}}
\vspace{1mm}
\end{minipage}
\begin{minipage}[c]{0.48\textwidth}
\includegraphics[width=\linewidth]{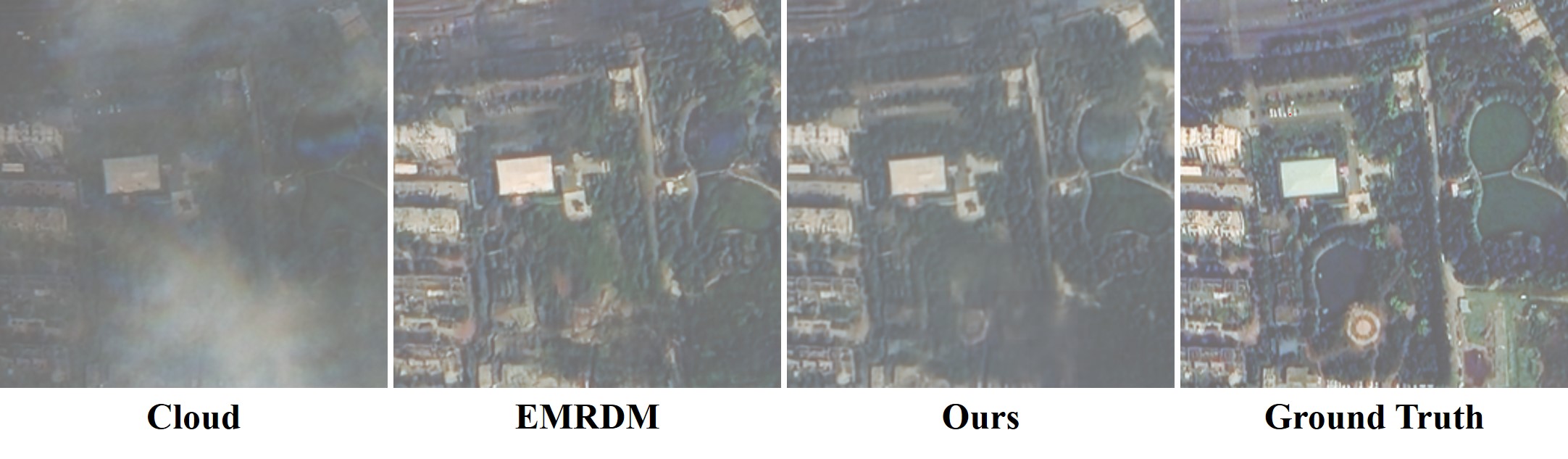}
\captionof{figure}{Quantitative evaluation results on cloud removal.}
\label{fig:cloud}
\end{minipage}
\end{table}

For the remote sensing cloud removal task, we conduct experiments using the CUHK-CR\cite{de-msda_memory} real-world public dataset, which includes the light cloud dataset CUHK-CR1\cite{de-msda_memory} and the heavy cloud dataset CUHK-CR2\cite{de-msda_memory}. Tab.\ref{tab:cloud} presents the quantitative comparison results for cloud removal, where our proposed SWFormer achieves the best PSNR and SSIM metrics, and the second-best LPIPS score. Notably, on the heavy cloud dataset CUHK-CR2, SWFormer shows a 1.14 dB PSNR improvement compared to the latest method, EMRDM\cite{emrdm}. Fig.\ref{fig:cloud} presents the qualitative comparison results for remote sensing cloud removal. Compared to EMRDM, our method achieves better fidelity in the restored cloud-free images, while EMRDM tends to introduce more artificial image artifacts, with more distortions or unrealistic details.

\subsection{Experimental Results on Model Complexity}
% \usepackage{multirow}
\begin{table*}[!b]
\setlength{\abovecaptionskip}{0cm}
\setlength{\belowcaptionskip}{0cm}
\renewcommand{\arraystretch}{1}
\centering
\caption{Experimental Results on Model Performance and Complexity Balance.\label{tab:overhead}}
\resizebox{\textwidth}{!}{
\setlength{\tabcolsep}{1mm}{
\begin{tabular}{cccccccccccccccccc}
\multicolumn{5}{c}{(a) Comparison on Deraining}                                                                   &  & \multicolumn{5}{c}{(b) Comparison on Low-light}                                                                  &  & \multicolumn{6}{c}{(c) Comparison on Deblurring}                                                                          \\ \cline{1-5} \cline{7-11} \cline{13-18} 
\multicolumn{1}{c|}{\multirow{2}{*}{Method}} & \multicolumn{2}{c|}{Deraining}      & \multicolumn{2}{c}{Overhead} &  & \multicolumn{1}{c|}{\multirow{2}{*}{Method}} & \multicolumn{2}{c|}{Low-light}     & \multicolumn{2}{c}{Overhead} &  & \multicolumn{1}{c|}{\multirow{2}{*}{Method}} & \multicolumn{3}{c|}{Defocus blur}           & \multicolumn{2}{c}{Overhead} \\ \cline{2-5} \cline{8-11} \cline{14-18} 
\multicolumn{1}{c|}{}                        & PSNR$\uparrow$  & \multicolumn{1}{c|}{SSIM$\uparrow$}   & \#Params         & \#FLOPs        &  & \multicolumn{1}{c|}{}                        & PSNR$\uparrow$  & \multicolumn{1}{c|}{SSIM$\uparrow$}  & \#Params         & \#FLOPs        &  & \multicolumn{1}{c|}{}                        & PSNR$\uparrow$   & SSIM$\uparrow$  & \multicolumn{1}{c|}{LPIPS$\downarrow$} & \#Params         & \#MACs         \\ \cline{1-5} \cline{7-11} \cline{13-18} 
\multicolumn{1}{c|}{(CVPR'22)Uformer\cite{Uformer}}        & 37.22 & \multicolumn{1}{c|}{0.9609} & 20.60M        & 41.09G       &  & \multicolumn{1}{c|}{(CVPR'22)Restormer\cite{Restormer}}      & 23.45 & \multicolumn{1}{c|}{0.841} & 26.10M        & 141.0G       &  & \multicolumn{1}{c|}{(CVPR'21)KPAC\cite{KPAC}}           & 27.082 & 0.816 & \multicolumn{1}{c|}{0.200} & 2.06M         & 98.5B        \\
\multicolumn{1}{c|}{(CVPR'22)Restormer\cite{Restormer}}      & 38.09 & \multicolumn{1}{c|}{0.9670} & 26.10M        & 141.0G       &  & \multicolumn{1}{c|}{(ECCV'22)LEDNet\cite{ledneteccv2022}}         & 24.16 & \multicolumn{1}{c|}{0.868} & 7.07M         & 35.92G       &  & \multicolumn{1}{c|}{(CVPR'21)IFANet\cite{ifanet}}         & 27.578 & 0.831 & \multicolumn{1}{c|}{0.186} & 10.48M        & 362.9B       \\
\multicolumn{1}{c|}{(TPAMI'22)IDT\cite{IDT}}           & 37.78 & \multicolumn{1}{c|}{0.9666} & 16.39M        & 58.44G       &  & \multicolumn{1}{c|}{(CVPR'22)SNR-Net\cite{snrnet}}        & 25.33 & \multicolumn{1}{c|}{0.880} & 4.01M         & 26.35G       &  & \multicolumn{1}{c|}{(NIPS'21)GKMNet\cite{gkmnet}}         & 27.275 & 0.828 & \multicolumn{1}{c|}{0.195} & 1.47M         & 147.8B       \\
\multicolumn{1}{c|}{(CVPR'23)DRSformer\cite{DRSformer}}      & 38.33 & \multicolumn{1}{c|}{0.9676} & 33.70M        & 242.9G       &  & \multicolumn{1}{c|}{(AAAI'23)LLFormer\cite{llformeraaai2023}}       & 24.62 & \multicolumn{1}{c|}{0.848} & 24.55M        & 22.52G       &  & \multicolumn{1}{c|}{(CVPR'22)DRBNet\cite{drbnet}}         & 27.869 & 0.838 & \multicolumn{1}{c|}{0.201} & 11.7M         & 346.6B       \\
\multicolumn{1}{c|}{(CVPR'24)NeRD-Rain\cite{nerd-rain}}      & 38.73 & \multicolumn{1}{c|}{0.9694} & 22.89M        & 156.3G       &  & \multicolumn{1}{c|}{(ICCV'23)RetinexFormer\cite{retinexformer}}  & 26.24 & \multicolumn{1}{c|}{0.877} & 1.61M         & 15.57G       &  & \multicolumn{1}{c|}{(CVPR'23)NRKNet\cite{NRKNet}}         & 28.295 & 0.847 & \multicolumn{1}{c|}{0.179} & 6.09M         & 552.8B       \\
\multicolumn{1}{c|}{(ECCV'24)FADformer\cite{fadformer}}      & 38.68 & \multicolumn{1}{c|}{0.9692} & 6.96M         & 48.51G       &  & \multicolumn{1}{c|}{(ECCV'24)SFHformer\cite{sfhformer}}      & 26.38 & \multicolumn{1}{c|}{0.894} & 1.04M         & 7.75G        &  & \multicolumn{1}{c|}{(ECCV'24)SFHformer\cite{sfhformer}}      & 28.371 & 0.847 & \multicolumn{1}{c|}{0.184} & 7.76M         & 267.5B       \\
\multicolumn{1}{c|}{(ECCV'24)SFHformer\cite{sfhformer}}      & 38.82 & \multicolumn{1}{c|}{0.9690} & 7.63M         & 50.59G       &  & \multicolumn{1}{c|}{(CVPR'25)CIDNet\cite{cidnetcvpr2025}}         & 26.35 & \multicolumn{1}{c|}{0.899} & 1.88M         & 7.57G        &  & \multicolumn{1}{c|}{(TPAMI'24)GGKMNet\cite{ggkmnet}}       & 28.412 & 0.848 & \multicolumn{1}{c|}{0.185} & 5.93M         & 518.9B       \\ \cline{1-5} \cline{7-11} \cline{13-18} 
\multicolumn{1}{c|}{(Ours)SWFormer-s}        & 38.57 & \multicolumn{1}{c|}{0.9680} & 0.95M         & 19.86G       &  & \multicolumn{1}{c|}{(Ours)SWFormer-s}        & 26.59 & \multicolumn{1}{c|}{0.889} & 1.22M         & 7.00G        &  & \multicolumn{1}{c|}{(Ours)SWFormer-s}        & 28.345 & 0.847 & \multicolumn{1}{c|}{0.181} & 3.10M         & 114.2B       \\
\multicolumn{1}{c|}{(Ours)SWFormer-m}        & 38.87 & \multicolumn{1}{c|}{0.9695} & 1.63M         & 27.65G       &  & \multicolumn{1}{c|}{(Ours)SWFormer-m}        & 26.92 & \multicolumn{1}{c|}{0.893} & 1.47M         & 9.41G        &  & \multicolumn{1}{c|}{(Ours)SWFormer-m}        & 28.569 & 0.855 & \multicolumn{1}{c|}{0.180} & 3.49M         & 145.2B       \\
\multicolumn{1}{c|}{(Ours)SWFormer-l}        & 39.23 & \multicolumn{1}{c|}{0.9716} & 1.91M         & 40.01G       &  & \multicolumn{1}{c|}{(Ours)SWFormer-l}        & 27.28 & \multicolumn{1}{c|}{0.906} & 1.54M         & 11.97G       &  & \multicolumn{1}{c|}{(Ours)SWFormer-l}        & 28.745 & 0.859 & \multicolumn{1}{c|}{0.177} & 3.60M         & 177.6B       \\ \cline{1-5} \cline{7-11} \cline{13-18} 
\end{tabular}}}
\end{table*}

In Tab.\ref{tab:overhead}, we compare model complexity (Params, FLOPs, MACs) and restoration performance (PSNR, SSIM, LPIPS) across the tasks of deraining, low-light enhancement and defocus deblurring. As shown in Tab.\ref{tab:overhead}(a,b,c), we present the average restoration performance across five deraining datasets (Rain200L\cite{RAIN200}, Rain200H\cite{RAIN200}, DDN-Data\cite{ddn}, DID-Data\cite{DID}, SPA-Data\cite{SPA-Data}), three low-light datasets (LOL-v1\cite{lolv1}, LOL-v2-real\cite{sparselolv2}, LOL-v2-syn\cite{sparselolv2}) and two defocus deblurring datasets (DPDD\cite{DPDD}, LFDOF\cite{aifnet}). Model complexity for deraining and low-light tasks is computed at 256×256 resolution, and for defocus deblurring, at 1280×720 resolution. The results demonstrate that our method achieves an optimal balance between performance and model complexity. Specifically, for rain removal, SWFormer-m uses only 21.4\% of the Params and 54.6\% of the FLOPs of SFHFormer\cite{sfhformer}, while outperforming it in both PSNR and SSIM. For defocus blur removal, SWFormer-m uses 58.8\% of the Params and 34.2\% of the MACs of GGKMNet\cite{ggkmnet}, maintaining an advantage in PSNR, SSIM and LPIPS.

\subsection{Experimental Results on Inference Latency}

In Tab.\ref{tab:time}, we compare model latency and restoration performance at two different resolutions. In Tab.\ref{tab:time}(a), using the low-resolution dataset Rain200L\cite{RAIN200} (400×600) at a 256×256 image resolution, SWFormer outperforms the latest methods in both restoration and latency. Notably, SWFormer-s achieves a 2.5$\times$ faster inference speed (25.9ms vs. 65.2ms) while improving PSNR by 0.16 dB over SFHFormer\cite{sfhformer}. In Tab.\ref{tab:time}(b), with the high-resolution dataset 4K-Rain13k\cite{udrmixer} (3840×2160) at 1024×1024 resolution, SWFormer-s achieves a 67$\times$ faster inference speed (40ms vs. 2682ms) compared to DRSformer\cite{DRSformer}, while delivering a 1.89 dB PSNR improvement.

% \usepackage{multirow}
\begin{table}[!h]
\setlength{\abovecaptionskip}{0cm}
\setlength{\belowcaptionskip}{0cm}
\renewcommand{\arraystretch}{1}
\centering
\caption{Experimental Results on Model Complexity.\label{tab:time}}
\resizebox{0.48\textwidth}{!}{
\setlength{\tabcolsep}{1mm}{
\begin{tabular}{ccccccccc}
\multicolumn{4}{c}{(a) Comparison on 256$\times$256}                                                &  & \multicolumn{4}{c}{(b) Comparison on 1024$\times$1024}                                              \\ \cline{1-4} \cline{6-9} 
\multicolumn{1}{c|}{\multirow{2}{*}{Method}} & \multicolumn{3}{c}{Rain200L}                  &  & \multicolumn{1}{c|}{\multirow{2}{*}{Method}} & \multicolumn{3}{c}{4K-Rain13k}                \\ \cline{2-4} \cline{7-9} 
\multicolumn{1}{c|}{}                        & PSNR$\uparrow$  & \multicolumn{1}{c|}{SSIM$\uparrow$}   & Latency &  & \multicolumn{1}{c|}{}                        & PSNR$\uparrow$  & \multicolumn{1}{c|}{SSIM$\uparrow$}   & Latency \\ \cline{1-4} \cline{6-9} 
\multicolumn{1}{c|}{(CVPR'22)Restormer\cite{Restormer}}      & 40.99 & \multicolumn{1}{c|}{0.9890} & 77.3ms  &  & \multicolumn{1}{c|}{(ICCV'21)SPDNet\cite{SPDNet}}         & 31.81 & \multicolumn{1}{c|}{0.9223} & 391ms   \\
\multicolumn{1}{c|}{(TPAMI'22)IDT\cite{IDT}}           & 40.74 & \multicolumn{1}{c|}{0.9884} & 74.8ms  &  & \multicolumn{1}{c|}{(CVPR'22)Restormer\cite{Restormer}}      & 33.02 & \multicolumn{1}{c|}{0.9335} & 1145ms  \\
\multicolumn{1}{c|}{(CVPR'23)DRSformer\cite{DRSformer}}      & 41.23 & \multicolumn{1}{c|}{0.9894} & 176.9ms &  & \multicolumn{1}{c|}{(TPAMI'22)IDT\cite{IDT}}           & 32.91 & \multicolumn{1}{c|}{0.9479} & 1935ms  \\
\multicolumn{1}{c|}{(CVPR'24)NeRD-Rain\cite{nerd-rain}}      & 41.71 & \multicolumn{1}{c|}{0.9903} & 115.8ms &  & \multicolumn{1}{c|}{(CVPR'23)DRSformer\cite{DRSformer}}      & 32.96 & \multicolumn{1}{c|}{0.9334} & 2682ms  \\
\multicolumn{1}{c|}{(ECCV'24)FADformer\cite{fadformer}}      & 41.80 & \multicolumn{1}{c|}{0.9906} & 72.1ms  &  & \multicolumn{1}{c|}{(ICCV'23)UDR-$\mathrm{S^2}$Former\cite{udrsformer}}    & 33.36 & \multicolumn{1}{c|}{0.9458} & 547ms   \\
\multicolumn{1}{c|}{(ECCV'24)SFHformer\cite{sfhformer}}      & 41.85 & \multicolumn{1}{c|}{0.9908} & 65.2ms  &  & \multicolumn{1}{c|}{(ArXiV'24)UDR-Mixer\cite{udrmixer}}     & 34.30 & \multicolumn{1}{c|}{0.9505} & 77ms    \\ \cline{1-4} \cline{6-9} 
\multicolumn{1}{c|}{(Ours)SWFormer-s}        & 42.01 & \multicolumn{1}{c|}{0.9908} & 25.9ms  &  & \multicolumn{1}{c|}{(Ours)SWFormer-s}        & 34.85 & \multicolumn{1}{c|}{0.9557} & 40ms    \\
\multicolumn{1}{c|}{(Ours)SWFormer-m}        & 42.23 & \multicolumn{1}{c|}{0.9913} & 37.4ms  &  & \multicolumn{1}{c|}{(Ours)SWFormer-m}        & 35.54 & \multicolumn{1}{c|}{0.9573} & 51ms    \\
\multicolumn{1}{c|}{(Ours)SWFormer-l}        & 42.70 & \multicolumn{1}{c|}{0.9922} & 55.4ms  &  & \multicolumn{1}{c|}{(Ours)SWFormer-l}        & 35.78 & \multicolumn{1}{c|}{0.9591} & 82ms    \\ \cline{1-4} \cline{6-9} 
\end{tabular}}}
\end{table}

\subsection{Ablation Studies}
To validate the effectiveness of our method, we conduct various ablation experiments on LMIMO and the individual components of SWFormer on the Rain200L\cite{RAIN200} dataset.

\begin{table}[h]
\setlength{\abovecaptionskip}{0cm}
\setlength{\belowcaptionskip}{0cm}
\renewcommand{\arraystretch}{1}
\centering
\caption{Ablation Study on LMIMO.\label{tab:ab_lmimo}}
\resizebox{0.48\textwidth}{!}{
\setlength{\tabcolsep}{1mm}{
\begin{tabular}{ccccccccc}
\hline
No. & SISO & MIMO & LMIMO & PSNR$\uparrow$ & SSIM$\uparrow$ & \#Params & \#FLOPs & Latency \\ \hline
1   &  \checkmark    &      &       &  42.11    &   0.9909   &   1.83M     &   38.48G &  52.784ms  \\
2   &      &   \checkmark   &       &   42.55   &   0.9919   &  1.90M      &    39.81G & 54.660ms  \\
3   &      &      &    \checkmark   &  42.70    &   0.9922   & 1.91M      &   40.01G   & 55.433ms \\ \hline
\end{tabular}}}
\end{table}
\subsubsection{Ablation Study about Lossless Multi-Input Multi-Output} Tab.\ref{tab:ab_lmimo} presents the comparisons under different inter-block structures (SISO, MIMO, LMIMO), showing that our LMIMO framework achieves superior performance. Specifically, compared to SISO, LMIMO achieves a 0.59dB PSNR improvement at the cost of 0.08M parameters, 1.53G computation and 2.649ms inference time. Compared to MIMO, LMIMO achieves a 0.15dB PSNR improvement with a cost of 0.01M parameters, 0.20G computation and 0.773ms inference time.

% \usepackage{multirow}
\begin{table}[h]
\setlength{\abovecaptionskip}{0cm}
\setlength{\belowcaptionskip}{0cm}
\renewcommand{\arraystretch}{1}
\centering
\caption{Ablation Study on Individual Components.\label{tab:ab_component}}
\resizebox{0.48\textwidth}{!}{
\setlength{\tabcolsep}{0.9mm}{
\begin{tabular}{c|c|ccc|ccc|cccc}
\hline
\multirow{10}{*}{\begin{tabular}[c]{@{}c@{}}Individual\\ Component\end{tabular}} & Spatial branch & \checkmark &            &            & \checkmark &            & \checkmark & \checkmark & \checkmark & \checkmark & \checkmark \\
                                                                                 & Wavelet branch &            & \checkmark &            & \checkmark & \checkmark &            & \checkmark & \checkmark & \checkmark & \checkmark \\
                                                                                 & Fourier branch &            &            & \checkmark &            & \checkmark & \checkmark & \checkmark & \checkmark & \checkmark & \checkmark \\
                                                                                 & Spatial loss   & \checkmark &            &            & \checkmark &            & \checkmark & \checkmark & \checkmark & \checkmark & \checkmark \\
                                                                                 & Wavelet loss   &            & \checkmark &            & \checkmark & \checkmark &            & \checkmark & \checkmark & \checkmark & \checkmark \\
                                                                                 & Fourier loss   &            &            & \checkmark &            & \checkmark & \checkmark & \checkmark & \checkmark & \checkmark & \checkmark \\ \cline{2-12} 
                                                                                 & DFFN\cite{Uformer}           &            &            &            &            &            &            & \checkmark &            &            &            \\
                                                                                 & GDFN\cite{Restormer}           &            &            &            &            &            &            &            & \checkmark &            &            \\
                                                                                 & MCFN\cite{sfhformer}           &            &            &            &            &            &            &            &            & \checkmark &            \\
                                                                                 & MSFN           & \checkmark & \checkmark & \checkmark & \checkmark & \checkmark & \checkmark &            &            &            & \checkmark \\ \hline
\multirow{2}{*}{Metric}                                                          & PSNR$\uparrow$           &       42.10     &      42.08      &     42.43       &    42.15        &       42.53     &   42.60         &      42.63      &      42.62      &      42.66      &    42.70        \\
                                                                                 & SSIM$\uparrow$           &      0.9910      &       0.9909     &    0.9918        &   0.9912         &      0.9919      &      0.9920      &     0.9921       &      0.9920       &      0.9921       &       0.9922     \\ \hline
\end{tabular}}}
\end{table}
\subsubsection{Ablation Study about Individual Components} Tab.\ref{tab:ab_component} presents the ablation results for SWFormer’s individual components, focusing on the Token Mixer and FFN at the intra-block level. To ensure fairness, we fine-tune each configuration to match model parameters and computational cost. For the Spatial-Wavelet-Fourier Mixer, we assess the contribution of each domain to feature extraction. The results reveal that multi-domain learning enhances feature extraction across multi-receptive fields, with the Fourier domain having the greatest impact, followed by the equally significant Spatial and Wavelet domains. In FFN, we compare our MSFN with MCFN\cite{sfhformer}, GDFN\cite{Restormer} and DFFN\cite{Uformer}, and find that MSFN yields better performance.

\begin{table}[h]
\setlength{\abovecaptionskip}{0cm}
\setlength{\belowcaptionskip}{0cm}
\renewcommand{\arraystretch}{1}
\centering
\caption{Ablation Study on Fourier Operation.\label{tab:ab_gate}}
\resizebox{0.48\textwidth}{!}{
\setlength{\tabcolsep}{1mm}{
\begin{tabular}{cccccccc}
\hline
No. & FDC\cite{sfhformer} & Gate Filter & PSNR$\uparrow$ & SSIM$\uparrow$ & \#Params & \#FLOPs & Latency \\ \hline
1   &   \checkmark  &             &  42.75    &   0.9923   &     1.89M   &   39.12G    &    73.181ms     \\ 
2   &  &  \checkmark   &   42.70   &   0.9922   &    1.91M    &   40.01G    &     55.433ms    \\ \hline
\end{tabular}}}
\end{table}
\subsubsection{Ablation Study about Fourier Operation} Tab.\ref{tab:ab_gate} compares the Gate Filter in the Fourier branch with the frequency dynamic convolution (FDC) from the conference version\cite{sfhformer}. While FDC achieves slightly better performance in PSNR, parameters and computational cost (0.05 dB, 0.02 M, 0.89 G), its inference time is significantly longer, being 1.32 times ours, with an additional 17.748 ms. Given this trade-off, we choose the Gate Filter, which offers a slight performance cost in exchange for a substantial reduction in latency.

\section{Conclusion}
In this work, we uncover a novel restoration prior from a Spatial-Wavelet-Fourier perspective, offering new insights for efficient and precise feature extraction. Guided by this, we propose SWFormer, an efficient Transformer-like backbone that integrates multi-domain learning to address image restoration. At the inter-block level, we introduce a Lossless Multi-Input Multi-Output framework, enabling dynamic outputs at varying sizes and restoration enhancement with minimal computational overhead. At the intra-block level, we retain the Transformer architecture but modify the Token Mixer and Feed-Forward Network. Specifically, in the Token Mixer, we replaces self-attention with a tri-branch structure for local-region-global modeling, reducing model complexity. In the FFN, we implement multi-scale learning to aggregate features across different domains and resolutions. We evaluate SWFormer through extensive experiments on 26 benchmarks across 10 restoration tasks, including dehazing, desnowing, rain streak/raindrop removal, cloud removal, motion blur, defocus blur, shadow removal, underwater enhancement and low-light enhancement. The results show that SWFormer achieves SOTA performance on most datasets, balancing performance, model size, computational cost and inference latency effectively.

% \section{Acknowledgments}
% This work is supported by the National Natural Science Foundation of China under Grant 62325101, Grant 62031001, and 62405014. This work is also supported by National Key Laboratory of Unmanned Aerial Vehicle Technology in NPU, (Grant No.WR202403), and the Fundamental Research Funds for the Central Universities.

% \begin{figure*}[t]
% 	\centering
% 	\includegraphics[width=\linewidth]{image/rain_result.jpg}
% 	\caption{The top .
% 	}
% 	\label{fig:intro}
% \end{figure*}

% \begin{figure*}[t]
% 	\centering
% 	\includegraphics[width=\linewidth]{image/raindrop_result_new.jpg}
% 	\caption{The top r.
% 	}
% 	\label{fig:intro}
% \end{figure*}

\bibliographystyle{splncs04}
\bibliography{main}

\begin{IEEEbiography}[{\includegraphics[width=1in,height=1.25in,clip,keepaspectratio]{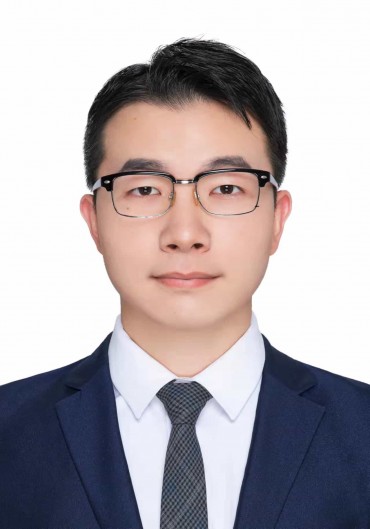}}]{Xingyu Jiang}
received the B.S. degree in aircraft
control and information engineering from the Image
Processing Center, School of Astronautics, Beihang
University, Beijing, China, in 2021, where he is currently working toward the Ph.D. degree in control
science and engineering.
His research interests include low-level computer vision, reinforcement learning and large language models.
\end{IEEEbiography}

\begin{IEEEbiography}[{\includegraphics[width=1in,height=1.25in,clip,keepaspectratio]{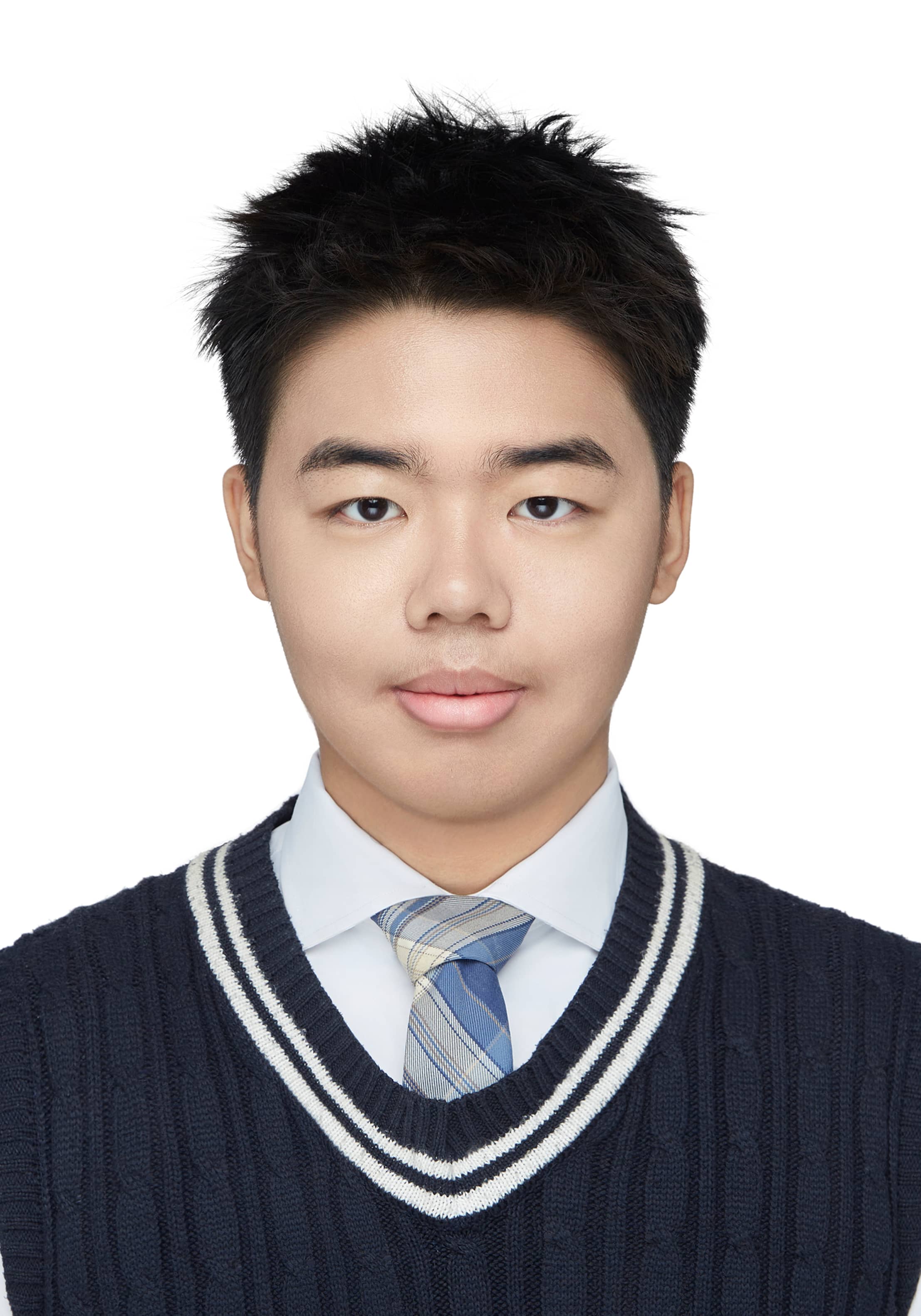}}]{Ning Gao}
received the B.S. degree in aircraft
control and information engineering from the Image
Processing Center, School of Astronautics, Beihang
University, Beijing, China, in 2024, where he is currently working toward the Master degree in control
science and engineering.
His research interests include machine learning , computer vision, reinforcement learning and large language models.
\end{IEEEbiography}

\begin{IEEEbiography}[{\includegraphics[width=1in,height=1.25in,clip,keepaspectratio]{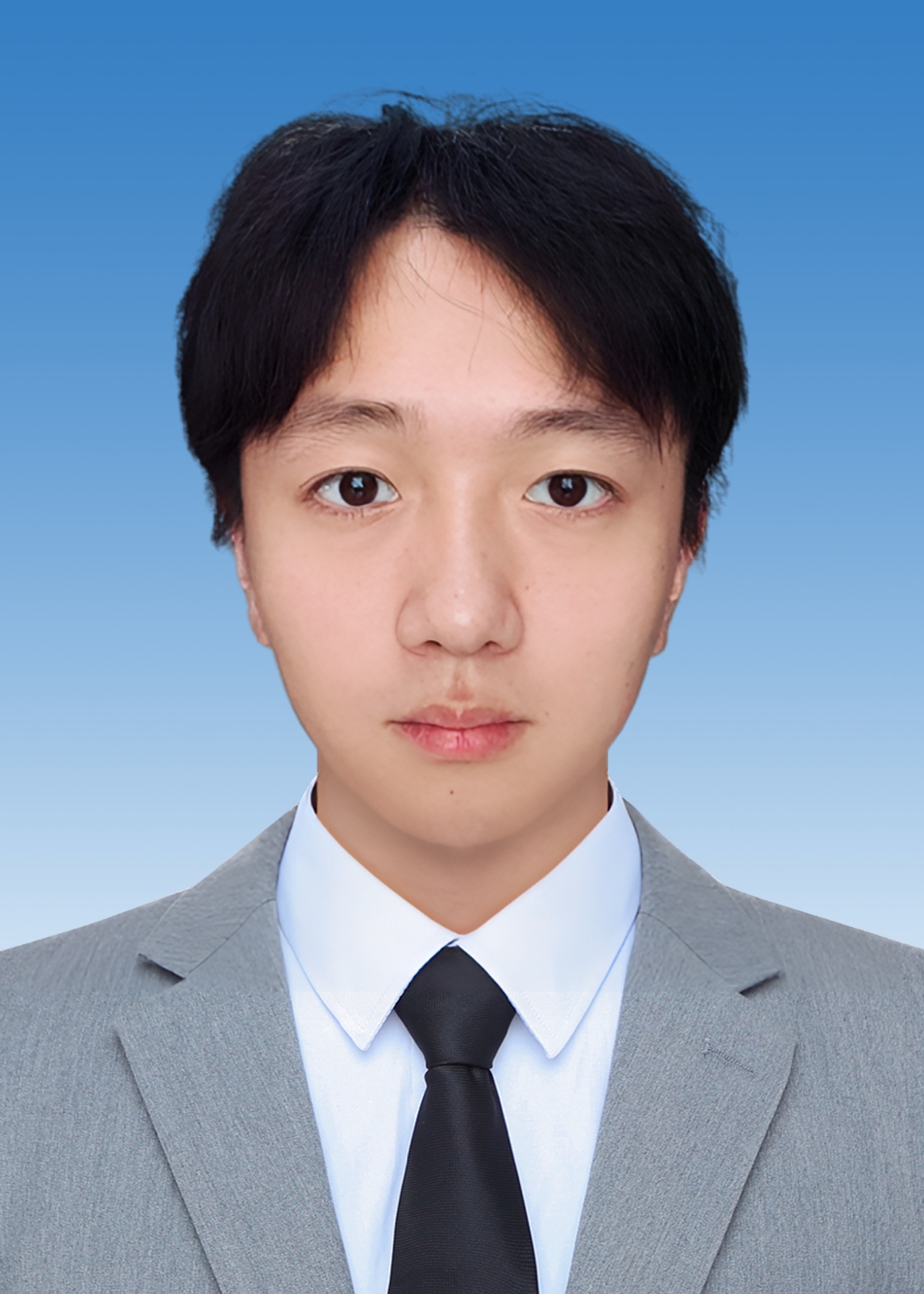}}]{Xiuhui Zhang}
received the B.S. degree in Guiding, Navigation and Control from School of Astronautics, Beihang
University, Beijing, China, in 2023. He is currently working toward the Ph.D. degree in artificial intelligence in Institute of Artificial Intelligence, Beihang
University, Beijing, China.
His research interests include machine learning and
its application in interdisciplinary domains.
\end{IEEEbiography}

\begin{IEEEbiography}[{\includegraphics[width=1in,height=1.25in,clip,keepaspectratio]{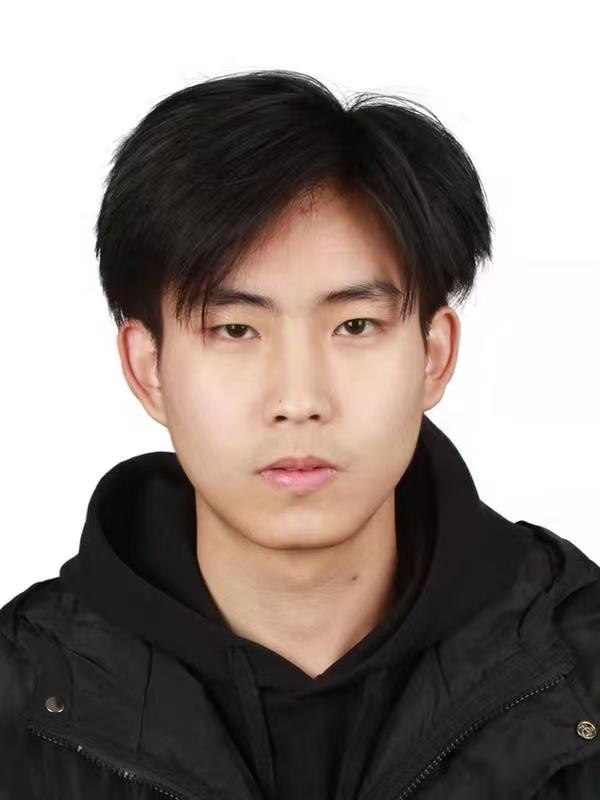}}]{Hongkun Dou}
received the B.S. degree in aircraft
control and information engineering from the Image
Processing Center, School of Astronautics, Beihang
University, Beijing, China, in 2021, where he is currently working toward the Ph.D. degree in control
science and engineering.
His research interests include machine learning and
its application in interdisciplinary domains.
\end{IEEEbiography}

\begin{IEEEbiography}[{\includegraphics[width=1in,height=1.25in,clip,keepaspectratio]{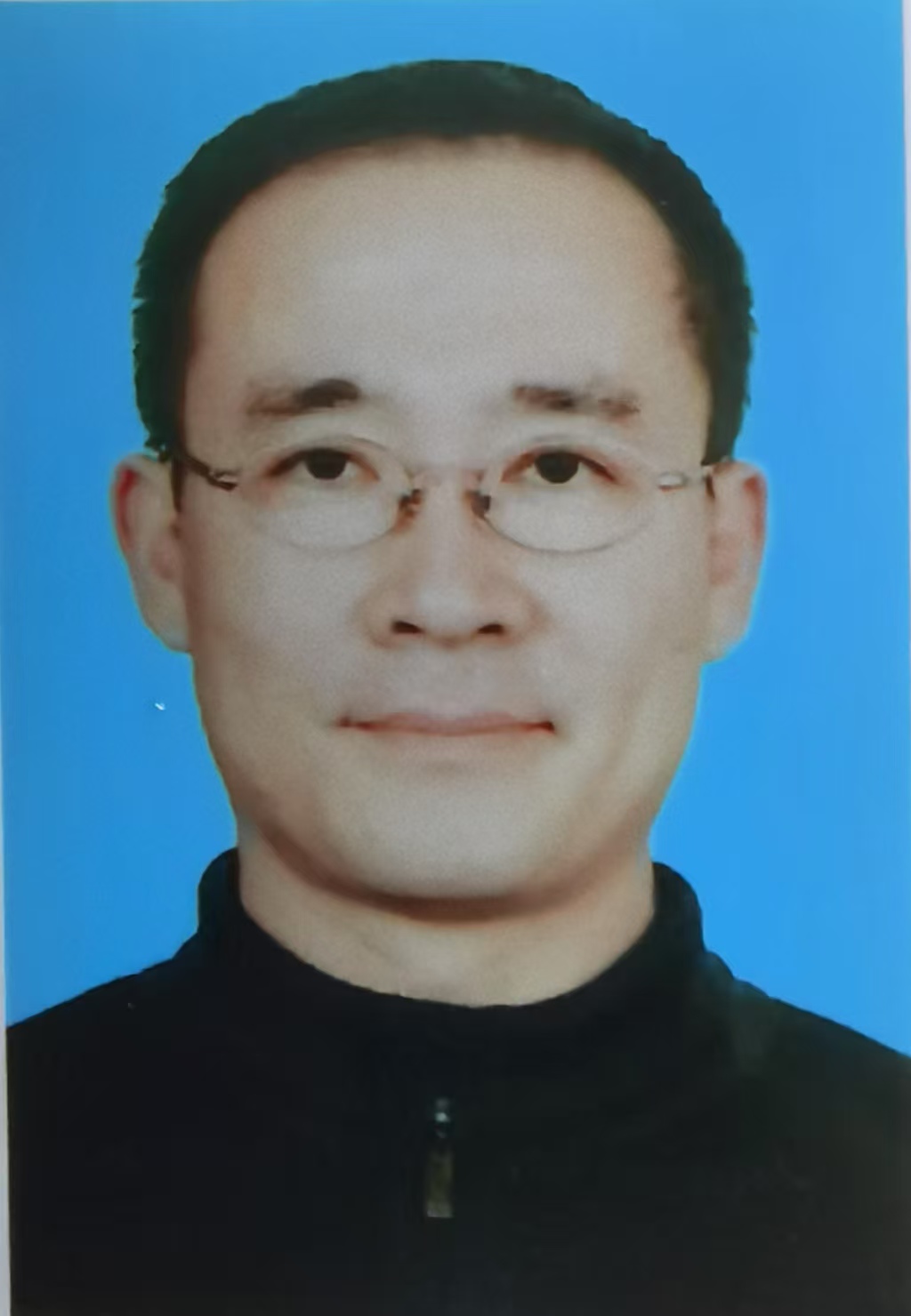}}]{Shaowen Fu} graduated from Harbin Institute of Technology in China in 2008, with Ph.D in control science and engineering. His research interests include control theory and navigation.
\end{IEEEbiography}

\begin{IEEEbiography}[{\includegraphics[width=1in,height=1.25in,clip,keepaspectratio]{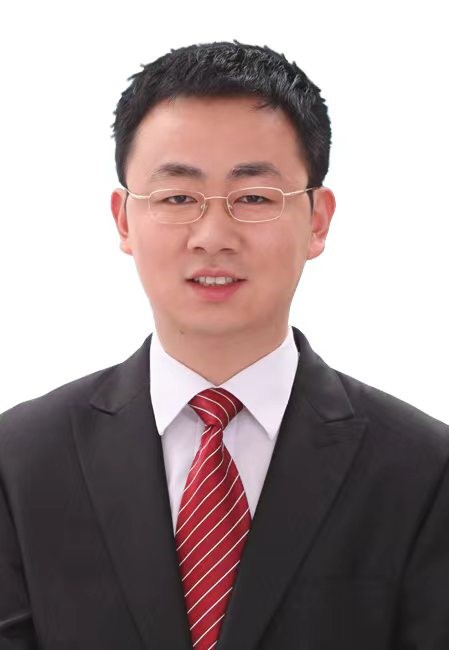}}]{Xiaoqing Zhong} received the B.S. and Ph.D. degrees in control science and technology from the Harbin Institute of Technology, Harbin, China, in 2005 and 2010, respectively. He is currently a Research Fellow of the China Academy of Space Technology, Beijing, China. His research interests include spacecraft system engineering and satellite communication.
\end{IEEEbiography}

\begin{IEEEbiography}[{\includegraphics[width=1in,height=1.25in,clip,keepaspectratio]{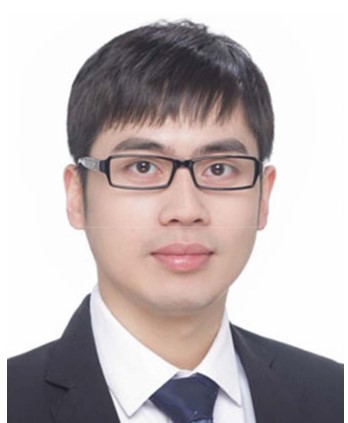}}]{Hongjue Li} received the Ph.D. degree in
 aerospace engineering fromBeihangUniversity,
 Beijing, China, in 2020.
 He is currently working as an associate professor with the School of Astronautics, Beihang
 University. His research interests include guid
ance and control, trajectory optimization, deep
 learning, and real-time optimal control with AI.
\end{IEEEbiography}

\begin{IEEEbiography}[{\includegraphics[width=1in,height=1.25in,clip,keepaspectratio]{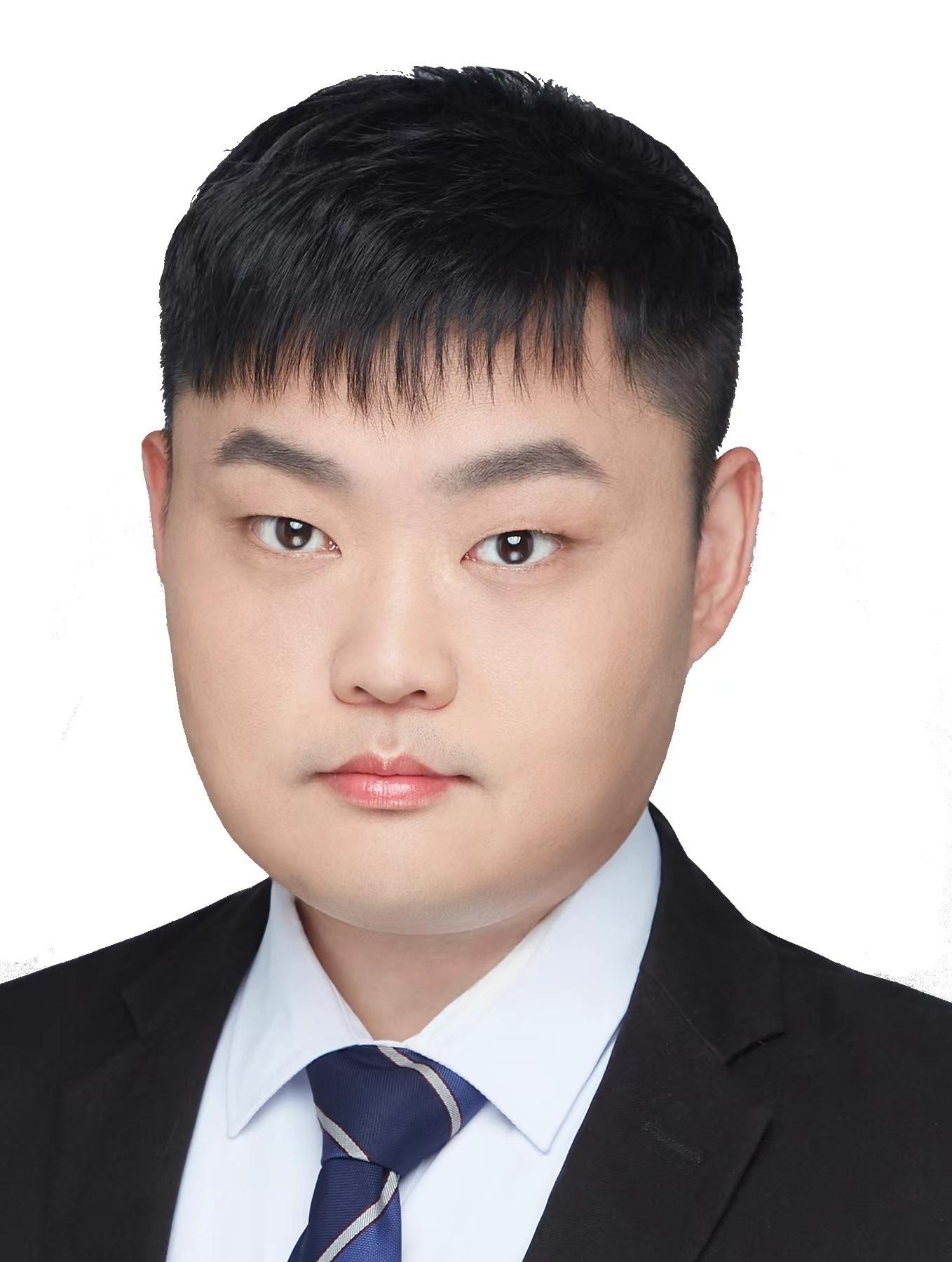}}]{Yue Deng}(Senior Member, IEEE) received the Ph.D.
degree in control science and engineering
from the Department of Automation, Tsinghua University, Beijing, China, in 2013.
He is currently a Professor with the School of
Astronautics, Beihang University, Beijing, China. His
research interests include machine learning, signal
processing, and computational biology.
He is an Associate Editor of \emph{IEEE Transactions on Neural Networks and Learning Systems} and \emph{IEEE Signal Processing Letters}.
\end{IEEEbiography}

\end{document}

\title{Supplementary Materials: Image Restoration via Multi-domain Learning}

\author{Xingyu Jiang, Ning Gao, Xiuhui Zhang, Hongkun Dou, Shaowen Fu, Xiaoqing Zhong, Hongjue Li and \\ Yue Deng,~\IEEEmembership{Senior Member,~IEEE,}}
\markboth{IEEE TRANSACTIONS ON PATTERN ANALYSIS AND MACHINE INTELLIGENCE}%
{JIANG \MakeLowercase{\textit{et al.}}: BILD$^{+}$: A Bi-level Framework for Image dehazing and Beyond}


\maketitle

\appendices

\begin{table*}[!b]
\setlength{\abovecaptionskip}{0cm}
\setlength{\belowcaptionskip}{0cm}
\renewcommand{\arraystretch}{1.4}
\centering
\caption{Details of the datasets for image restoration tasks.\label{tab:datasets}}
\resizebox{\textwidth}{!}{
\setlength{\tabcolsep}{1.2mm}{
\begin{tabular}{cccccc}
\hline
Task                                    & Dataset     & Average Resolution & Train Number & Test Number & Testset Rename \\ \hline
\multirow{5}{*}{Dehazing}               & RESIDE-ITS\cite{reside}  & $620\times460$     & 13990        & 500         & SOTS-Indoor    \\
                                        & RESIDE-OTS\cite{reside}  & $550\times432$     & 313950       & 500         & SOTS-Outdoor   \\
                                        & O-HAZE\cite{Ohaze}      & $800\times800$     & 14490        & 2070        & O-HAZE         \\
                                        & NH-HAZE\cite{nhHAZE}     & $800\times800$     & 9000         & 1000        & NH-HAZE        \\
                                        & DENSE-HAZE\cite{denseHAZE}  & $800\times800$     & 9000         & 1000        & DENSE-HAZE     \\ \hline
\multirow{8}{*}{Deraining}              & Rain200L\cite{RAIN200}    & $480\times320$     & 1800         & 200         & Rain200L       \\
                                        & Rain200H\cite{RAIN200}    & $480\times320$     & 1800         & 200         & Rain200H       \\
                                        & DID-Data\cite{DID}    & $512\times512$     & 12000        & 1200        & DID-Data       \\
                                        & DDN-Data\cite{ddn}    & $512\times384$     & 12600        & 1400        & DDN-Data       \\
                                        & SPA-Data\cite{SPA-Data}    & $512\times512$     & 638492       & 1000        & SPA-Data       \\
                                        & Raindrop\cite{AttentiveGANraindrop}    & $720\times480$     & 861          & 58          & Raindrop-A     \\
                                        & Raindrop\cite{AttentiveGANraindrop}    & $720\times480$     & 0            & 239         & Raindrop-B     \\
                                        & 4K-Rain13k\cite{udrmixer}  & $3840\times2160$   & 12500        & 500         & 4K-Rain13k     \\ \hline
\multirow{2}{*}{Motion Deblurring}      & GoPro\cite{GOPRO}       & $1280\times720$    & 2103         & 1111        & GoPro          \\
                                        & HIDE\cite{HIDE}        & $1280\times720$    & 0            & 2025        & HIDE           \\ \hline
\multirow{2}{*}{Defocus Deblurring}     & DPDD\cite{DPDD}        & $1120\times1680$   & 350          & 76          & DPDD           \\
                                        & LFDOF\cite{aifnet}       & $688\times1008$    & 11261        & 725         & LFDOF          \\ \hline
\multirow{3}{*}{Desnowing}              & CSD\cite{HDCW-NetCSD}         & $640\times480$     & 8000         & 2000        & CSD(2000)      \\
                                        & SRRS\cite{JSTASRSRRS}        & $640\times480$     & 15000        & 2000        & SRRS(2000)     \\
                                        & Snow100K\cite{desnownesnow100k}    & $640\times420$     & 50000        & 2000        & Snow100K(2000) \\ \hline
\multirow{2}{*}{Underwater Enhancement} & UIEB\cite{waternetuieb}        & $860\times590$     & 750          & 90          & U-90           \\
                                        & LSUI\cite{U-shapeTranslsui}        & $446\times310$     & 3500         & 400         & L-400          \\ \hline
\multirow{3}{*}{Low-light Enhancement}  & LOL-v1\cite{lolv1}      & $600\times400$     & 485          & 15          & LOL-v1         \\
                                        & LOL-v2-real\cite{sparselolv2} & $600\times400$     & 689          & 100         & LOL-v2-real    \\
                                        & LOL-v2-syn\cite{sparselolv2}  & $384\times384$     & 900          & 100         & LOL-v2-syn     \\ \hline
Shadow Removal                          & AISTD\cite{aistd}       & $640\times480$     & 1330         & 540         & AISTD          \\ \hline
\multirow{2}{*}{Cloud Removal}          & CUHK-CR1\cite{de-msda_memory}    & $512\times512$     & 534          & 134         & CUHK-CR1       \\
                                        & CUHK-CR2\cite{de-msda_memory}    & $512\times512$     & 448          & 111         & CUHK-CR2       \\ \hline
\end{tabular}}}
\end{table*}
\section{Datasets and Experimental Details}
In tab.\ref{tab:datasets}, we list various image restoration datasets (average resolutions and benchmark scales) adopted for training and evaluation. Next, we describe them within each individual image restoration task for implementation details.

\subsection{Image Dehazing}
We conduct dehazing experiments on both the synthetic benchmark RESIDE\cite{reside} and real-world hazy datasets, including Dense-Haze\cite{denseHAZE}, O-Haze\cite{Ohaze} and NH-Haze\cite{nhHAZE}. For the RESIDE dataset, we train our model separately for indoor and outdoor scenarios, then test on the corresponding SOTS dataset. Specifically, for the indoor experiment, ITS contains 13,990 hazy/clear pairs for training, while SOTS-indoor consists of 500 hazy/clear pairs for testing. For the outdoor experiment, OTS contains 313,950 hazy/clear pairs for training, and SOTS-outdoor includes 500 hazy/clear pairs for testing. Our SWFormer model is trained for 800k steps on both ITS and OTS with a batch size of 16. For the real-world datasets, O-HAZE, NH-HAZE and DENSE-HAZE with resolutions of 3600×3000, 1600×1200 and 1600×1200 respectively, we apply crop operation to each image for training. For O-HAZE, we use a stride of 128 to crop each image to a size of 800×800, resulting in a training set of 14,490 and a testing set of 2,070 images. For NH-HAZE and DENSE-HAZE, we use a stride of 40 to crop each image to 800×800, yielding a training set of 9,000 and a testing set of 1,000 images for each dataset. For these three datasets, SWFormer is trained for 10k steps with a patch size of 800×800.

\subsection{Image Deraining}
Following previous work\cite{DRSformer}, we evaluate PSNR and SSIM on the Y channel in the YCbCr color space. Our experiments are conducted on several datasets, including Rain200H\cite{RAIN200}, Rain200L\cite{RAIN200}, DID-Data\cite{DID}, DDN-Data\cite{ddn}, SPA-Data\cite{SPA-Data} and 4K-Rain13k\cite{udrmixer}. Specifically, Rain200H and Rain200L contain 1,800 synthetic rainy/clear image pairs for training and 200 for testing. DID-Data and DDN-Data consist of 12,000 and 12,600 synthetic rainy/clear pairs, respectively, with varying rain directions and density levels, and testing sets of 1,200 and 1,400 pairs. SPA-Data is a large-scale real-world dataset comprising 638,492 rainy/clear pairs for training and 1,000 for testing. 4K-Rain13k is a high-resolution dataset (3840×2160) containing 12,500 rainy/clear image pairs for training and 500 for testing. SWFormer is trained for 600k steps across all these datasets.

\subsection{Image Raindrop Removal}
Following previous work\cite{MAXIM}, We perform experiments on Raindrop\cite{AttentiveGANraindrop} for image raindrop removal. The Raindrop dataset contains 861 raindrop/clear pairs for training and, 58 ones of testset A and 239 ones of testset B for evaluation, respectively. SWFormer is trained for 60k steps.

\subsection{Image Motion Deblurring}
We evaluate SWFormer on GoPro\cite{GOPRO} and HIDE\cite{HIDE} for single-image motion deblurring, following recent methods \cite{Restormer, sfnet}. Gopro dataset contains 2103 blurry/clear training pairs and 1111 blurry/clear testing ones, which is obtained by a high-speed camera. To assess the robustness and generalizability of our approach, we conduct an evaluation by deploying the model trained on GoPro dataset directly onto the HIDE dataset. The HIDE dataset comprises 2025 pairs of blurry/clear images specifically curated for evaluation. SWFormer is trained for 800k steps with a batch size of 12.

\subsection{Image Defocus Deblurring}
We compare our proposed SWFormer with state-of-the-art methods on the DPDD\cite{DPDD} and LFDOF\cite{aifnet} datasets for single-image defocus deblurring. DPDD consists of images from 500 distinct indoor and outdoor scenes, with each scene containing four images: right-view, left-view, center-view, and the associated all-in-focus ground truth image. The DPDD dataset is split into training, validation and testing sets, comprising 350, 74, and 76 scenes (37 indoor and 39 outdoor), respectively. For our experiments, we train the model on the center-view image along with its corresponding ground truth. LFDOF is a large-scale defocus deblurring dataset with 11,261 blurry/clear image pairs for training and 725 pairs for testing. SWFormer is trained on both DPDD and LFDOF datasets for 150k steps.

\subsection{Image Desnowing}
We compare our method on CSD\cite{HDCW-NetCSD}, SRRS\cite{JSTASRSRRS} and snow100K\cite{desnownesnow100k} dataset with existing state-of-the-art methods for image desnowing. CSD is a large-scale snow dataset consisting of 8000 synthesized snow images. SRRS contains 15000 synthesized snow images and Snow100K has 100k synthesized snowy images. The dataset settings follow previous works\cite{focalnet}, where we randomly sample 2500 image pairs from the training set for training and 2000 images from testing set for evaluation. SWFormer is trained for 300k steps on each dataset.

\subsection{Image Underwater Enhancement}
We compare our method on UIEB\cite{waternetuieb} and LSUI\cite{U-shapeTranslsui} datasets with existing state-of-the-art methods for underwater image enhancement. The UIEB dataset contains 890 real underwater images with corresponding ground truths. We randomly selected 750 pairs for training, 50 pairs for validation and 90 pairs for testing (U-90). LSUI, which builds in a similar method to UIEB but its scale is larger, contains 4279 image pairs. We randomly selected 3500 pairs for training, 379 pairs for validation and 400 pairs for testing (L-400). SWFormer is trained for 60k and 200k steps on UIEB and LSUI, respectively.

\subsection{Image Low-light Enhancement}
We evaluate SWFormer on LOL-v1\cite{lolv1} and LOL-v2\cite{sparselolv2} for low-light image enhancement, following recent methods\cite{retinexformer}. The LOL dataset comprises versions v1 and v2. LOL-v2 is further categorized into real and synthetic subsets. The division of training and testing sets follows a ratio of 485:15, 689:100, and 900:100 for LOL-v1, LOL-v2-real, and LOL-v2-synthetic, respectively. SWFormer is trained for 100k steps on LOL-v1, LOL-v2-real and LOL-v2-syn with a patch size of 256 $\times$256, 256 $\times$256 and 128 $\times$ 128, respectively.

\subsection{Image Shadow Removal}
We conduct experiments on AISTD\cite{aistd} shadow removal dataset. AISTD is an adjusted version of ISTD\cite{istd}, which consists of 1330 shaodw/clear/mask triplets for training and 540 triplets for testing. We follow the previous method and use ground truth masks during the training. As for evaluation, we calculate the difference between the shadow area and the whole image via the masks. SWFormer is trained for 150k steps.

\subsection{Image Cloud Removal}
Following latest work\cite{emrdm}, we adopt the CUHK-CR1\cite{de-msda_memory} and CUHK-CR2\cite{de-msda_memory} datasets for remote sensing cloud removal, which consist of images captured by the Jilin-1 satellite with a size of 512×512. CUHK-CR1 contains 668 images of thin clouds, while CUHK-CR2 includes 559 images of thick clouds. These two datasets collectively form the CUHK-CR dataset. With an ultra-high spatial resolution of 0.5 m, the images encompass four bands: RGB and near-infrared (NIR). Following \cite{de-msda_memory}, the CUHK-CR1 dataset is split into 534 training and 134 testing images, while CUHK-CR2 is divided into 448 training and 111 testing images. SWFormer is trained for 150k steps on both CUHK-CR1 and CUHK-CR2 datasets.

%%%%%%%%% REFERENCES
\bibliographystyle{splncs04}
\bibliography{main}
% \bibliographystyle{ieeetr}
% \bibliography{egbib}

\clearpage